\documentclass[11pt,a4paper]{article}
\pdfoutput=1
\usepackage{pgf}
\usepackage{tikz}
\usepackage{graphicx}
\usetikzlibrary{arrows,automata}
\usepackage{amssymb,amsmath,amsfonts}
\usepackage{longtable}
\usepackage{verbatim}
\usepackage{color}
\usepackage[mathscr]{euscript}
\usepackage{framed}
\usepackage{mdframed}
\usepackage{amsmath}
\usepackage{amsfonts}
\usepackage{mathrsfs}
\usepackage{amssymb}
\usepackage{mathrsfs}  
\usepackage{subcaption}
\usepackage{caption}
\usepackage[many]{tcolorbox}
\usepackage{appendix}

\usepackage[left=26mm,top=20mm,right=26mm,bottom=20mm]{geometry}

\usepackage{color}
\usepackage{cite}
\usepackage{hyperref}
\hypersetup{colorlinks, citecolor=bluscuro, linkcolor=black, urlcolor=bluscuro}
\definecolor{rossos}{cmyk}{0,1,1,0.55}
\definecolor{bluscuro}{rgb}{0.15, 0.2, .85}
\definecolor{bluchiaro}{cmyk}{1,.3,0.,0.1}

\graphicspath{{./Figures/}}

\newcommand{\be}{\begin{equation}}
\newcommand{\ee}{\end{equation}}
\newcommand{\bea}{\begin{eqnarray}}
\newcommand{\eea}{\end{eqnarray}}
\newcommand{\beq}{\begin{equation}}
\newcommand{\eeq}{\end{equation}}

\begin{document}

\vspace{0.1in}

\begin{center}
{\Large\bf\color{black}
Reheating in runaway inflation models  via the evaporation of  mini primordial black holes
}\\
\bigskip\color{black}
\vspace{.5cm}
{ {\large  {Ioannis Dalianis}$^1$ and  {George P. Kodaxis}$^2$}
\vspace{0.3cm}
} \\[5mm]
{\it {Department of Physics, University of Athens, University Campus, Zographou 157 84, Greece}}
\\[2mm]
$^1 \text{idalianis@phys.uoa.gr}$,  $^2 \text{gekontax@phys.uoa.gr}$
\\[2mm]

\end{center}

\vskip.2in

\noindent
\rule{16.6cm}{0.4pt}

\vspace{.3cm}
\noindent
{\bf \large {Abstract}}
\vskip.15in 
We investigate the cosmology of mini Primordial Black Holes (PBHs) produced by large density perturbations that collapse during a stiff fluid domination phase. Such a phase can be realized by a runaway-inflaton model that crosses an inflection point or a sharp feature at the last stage  of inflation.
Mini PBHs evaporate promptly  and reheat the early universe.
In addition, we examine two notable implications of this scenario: the possible presence of PBH evaporation remnants in galaxies and 
a non-zero residual potential energy density for the runaway inflaton that might play the role of  the dark energy.
We specify the parameter space that this scenario can be realized 
and we find that a transit PBH domination phase is necessary due to gravitational wave (GW) constraints. 
A distinct prediction of the scenario is a compound GW signal that might be probed by current and future  experiments.
 We also demonstrate our results employing   an explicit  inflation model.

\noindent

\bigskip

\noindent
\rule{16.6cm}{0.4pt}

\vskip.4in

\tableofcontents

\section{Introduction}
The detection of gravitational waves (GWs) from black hole mergers  has confirmed that black holes are undoubtedly present in our universe with a notable abundance and a quite extended mass range. 
Black holes with masses from about 3 solar masses ($M_\odot$) up to 160$M_\odot$ have been reported \cite{LIGOScientific:2018mvr}. 
The black hole catalog is extended by the electromagnetic observations of supermassive black holes in the center of galaxies with masses that reach tens of billions of solar masses. Apparently massive black holes are easier to detect.
Nevertheless, it is natural to ask whether black holes  spread  across the mass spectrum  towards smaller values as well. 

From simple theoretical considerations,  the minimum  black hole mass is of the order of the Planck mass $m_\text{Pl}=G^{-1/2} \sim 10^{-5}$ g. Upper bound does not exist; the entire universe energy density could in principle collapse into a black hole. What is of special interest in the small mass limit, in sharp contrast to the large mass limit,  is that the predicted  Hawking radiation is manifest \cite{Hawking:1975vcx, Hawking:1974rv}. 
The emitted thermal radiation has a temperature $T_\text{BH}\propto 1/M_\text{BH}$ and black holes with mass $M_\text{BH}$ less than $10^{15}$ grams have a lifetime less than the age of the universe and are not expected to wander around in space.
Moreover, black holes with  such a small mass, far below the Chandrasekhar limit \cite{Chandrasekhar:1931ih} of 1.4$M_\odot$, cannot be produced by stellar collapse processes. However, small mass black holes can be produced from the collapse of overdense regions in the highly dense early universe  \cite{Carr:1974nx, Carr:1975qj}. 

Black holes of small mass are therefore relevant for the early universe cosmology.
They are produced in fractions of the  first second of the cosmic evolution and evaporate promptly.
 Indeed, black holes with mass less than $10^9$ grams evaporate before big bang nucleosynthesis (hereafter referred as BBN) and leave no apparent observational signature behind \cite{Carr:2020gox, Anantua:2008am}. 
 The  fact that mini black holes transform nearly their entire mass into thermal radiation is a mechanism that contributes to the entropy observed in our universe.  
Actually,  mini black holes can reheat the universe if no other radiation sources are efficient enough.
 This is the central topic of this paper.
 
 In the conventional inflationary scenario, reheating is understood as the decay of the inflaton field itself into entropic degrees of freedom. This is achieved either via parametric resonance or by single particle decays~\cite{Abbott:1982hn, Kofman:1997yn}. However, a stage of an oscillatory phase for the inflaton field is necessary and this is not always the case.
 There are models where the inflaton potential $V(\phi)$ gradually decreases after the end of the inflationary phase or even does not  have a minimum at  all. These are the so-called runaway models. This class of post-inflationary models has also been dubbed  ``NO-models” due to their  non-oscillatory behavior \cite{Felder:1999pv}. Reheating in these models seems problematic because it is generally too  feeble. The main source of radiation is the gravitational particle production, \cite{Ford:1986sy} which, in its original version,  fails to reheat the universe successfully.  Reheating can become efficient if the inflaton  sector is augmented with a tailor made coupling with an extra scalar field that depletes the inflaton energy density \cite{Felder:1998vq, Felder:1999pv} via instant preheating. A third post-inflationary reheating mechanism is via the primordial black hole production and evaporation, which 
 has been first discussed in Ref. \cite{Barrow:1990he, Carr:1994ar}

Inflationary models without a minimum are predicted in many theories beyond the standard model of particle physics. Some notable early examples can be found  in stringy set ups,  supergravity, braneworlds  as well as in many phenomenological models, see~\mbox{\cite{Townsend:2001ea, Copeland:1997et, Copeland:2000hn,  Kolda:2001ex, Elizalde:2004mq, Kehagias:2004bd, Russo:2004ym, Dalianis:2014nwa}}, just to mention a few. 
An additional phenomenological motivation to study these models is the  unification they offer in describing the two accelerating stages of  our universe: inflation and dark energy.
In fact, runaway models have been  widely   introduced  to describe quintessence fields. This unified framework has been called quintessential inflation  \cite{Peebles:1998qn, Spokoiny:1993kt}, and recently notable progress on this direction has taken place \cite{Dimopoulos:2017zvq, Dimopoulos:2017tud, Akrami:2017cir, Dimopoulos:2019gpz, Dimopoulos:2020pas}.

 In this work, we consider runaway inflationary models and implement the reheating of the universe via the Hawking evaporation of primordial black holes produced by the collapse of overdense regions of the inflaton field itself.
Actually, a kination phase due to a stiff energy component, $w>1/3$, can  be realized only at sufficiently early times of the cosmic evolution, where $w$ is the equation of state.
A simplistic comparison of the different red-shifts $a^{-3(1+w)}$ of the various  energy components shows that the scenario of a  stiff fluid domination at the early universe and a cosmological constant domination at the late universe is a plausible guess.   
Furthermore, since the inflaton potential energy density does not vanish, we contemplate the scenario that  the dark energy density of the universe is the present-day energy density of the inflaton.  
We employ a particular inflation model,
built in the framework of $\alpha$-attractors \cite{Kallosh:2013yoa, Dalianis:2018frf}.
Notably, the fact that the generated PBHs are ultra light implies that the  $(n_s, r)$ inflationary predictions lay in the sweet spot region of Planck and BICEP~\cite{Akrami:2018odb, BICEP:2021xfz}. This scenario, which is  remarkably economic in terms of ingredients,
  has been introduced and analyzed in Ref.  \cite{Dalianis:2019asr}.

 We specify the conditions under which this scenario can be realized extending the work  of Ref.  \cite{Dalianis:2019asr}.   We take into consideration the impact of primary and secondary GWs on the BBN and CMB observables and we constrain the ($M_\text{PBH}, \beta$) parameter space. We find out that a transit PBH domination phase, taking place between kination and radiation domination, is necessary.  
Interestingly enough, the allowed parameter space can become even more narrow  by near future GW probes. Last, but not least, an additional striking implication of the very same scenario is that the 
 evaporation of the mini black holes might leave a stable remnant behind \cite{Coleman:1991ku, Barrow:1992hq, Carr:1994ar,  Adler:2001vs, Suranyi:2010yt, Alexander:2007gj, Scardigli:2010gm, Torres:2013kda, Chen:2014jwq, Dymnikova:2015yma, Lennon:2017tqq, Raidal:2018eoo, Rasanen:2018fom,  Nakama:2018utx, Morrison:2018xla, Kovacik:2021qms}. 
  We notice that remnants must have a mass in a particular range and in correlation with the PBH mass in order to have a cosmologically significant abundance.

The paper is organized as follows. In the next section we discuss the process of PBH formation during a kination era. In Section \ref{S3}
 we motivate the formation of ultra-light PBHs and we review the basic elements of PBH evaporation and PBH remnants. In Section~\ref{S4}, we study the evolution of the energy densities of the inflaton, PBHs and remnants and we specify the relevant for our discussion  cosmological quantities. 
In Section \ref{S5}, we examine the primary and secondary GWs and their impact on BBN and CMB. That section contains the main result of this paper, since we derive the main constraints on the PBH evaporation reheating of the runaway inflation models. In Section \ref{S6}, we introduce an explicit runaway inflation model that features an inflection point, and in Section \ref{S7} we discuss the quintessential inflation scenario and study the post-inflationary evolution of the inflaton field. Finally, in Section \ref{S8} we conclude.  The alternative to this  scenario mechanisms, gravitational reheating and instant (p)reheating, are reviewed in Appendix \ref{A1}.

\section{PBH Formation during a Kination Era}\label{S2}

Let us assume that at the early moment $t_\text{form}$
a fraction $\beta $ of the energy density of the universe collapses and forms PBHs. 
The mass density of the PBHs is 
\begin{align} \label{rhobeta}
\rho_\text{PBH} = \gamma \beta \rho_\text{tot}
\end{align}
 at the moment of formation, where $\rho_\text{tot}=3H^2M^2_\text{Pl}$ with $M_\text{Pl}=m_\text{Pl}/\sqrt{8\pi}$, the reduced Planck mass. 
 The $\gamma$ parameter is the fraction of the collapsing mass that finds itself inside the black hole.
 If the collapsing region is the Hubble horizon, the mass of the black hole is $M_\text{PBH}=\gamma M_\text{hor}= \gamma(4/3)\pi \rho_\text{tot}  H^{-3}$, where $M_\text{hor}=\frac34 (1+w) m^2_\text{Pl} t$, $w$ the equation of state of the background fluid and $t$ the cosmic time.
The formation probability is usually rather small,  $\beta\ll 1$, and  the background energy density $\rho_\text{bck}=(1-\gamma\beta)\rho_\text{tot}$ is approximately equal to  $\rho_\text{tot}$.
The PBHs are non-relativistic matter and their number density,
\begin{align}
n_\text{PBH}=\frac{\rho_\text{PBH}}{M_\text{PBH}}
\end{align}
 scales like  $a^{-3}$, while 
the background energy density scales as $\rho \propto a^{-3(1+w)}$. 
If PBHs have an extended mass spectrum one writes
$n_\text{PBH}(M)={d N_\text{PBH}(M)}/{d\ln M}$,
where $N_\text{PBH}$ is the integrated  PBH number density  up to  mass $M$.

Let us  assume that the bulk energy density is in the form of stiff fluid,
realized by the domination of the kinetic energy of a scalar field.
Such a fluid has a barotropic parameter $w\simeq 1$ and the phase is called kination (KIN).
A non-oscillatory inflaton field can give rise to a kination phase. 
During kination the scale factor goes like $a(t)\propto t^{1/3}$, 
the Hubble horizon mass is 
$M_\text{hor}=(3/2)m^2_\text{Pl} t$, and the scaling of the background energy density 
 redshifts like $\rho_\text{KIN} \propto a^{-6}$.

\begin{figure}[!htbp]
  \includegraphics[width=.496\linewidth]{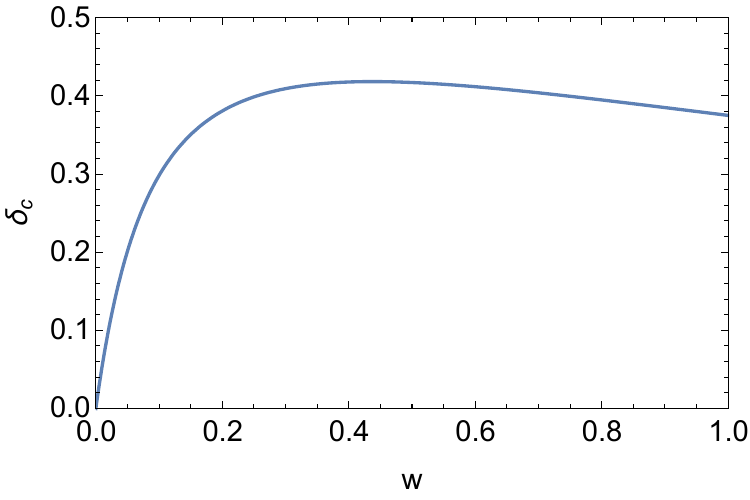}
  \includegraphics[width=.49\linewidth]{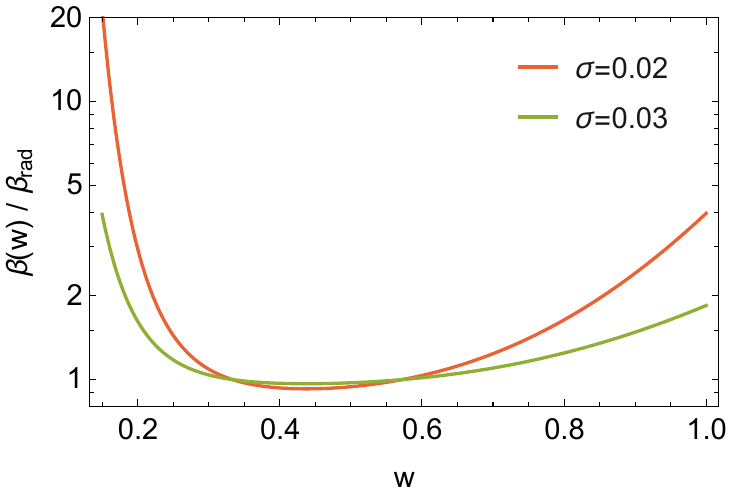}
\caption{In the \textbf{left panel} the dependence of the threshold value  (\ref{dc}) on the EOS, $w$, is plotted. In the \textbf{right panel} the increase of $\beta$, and consequently the PBH number density, with the hardening of the EOS is shown. The $\beta(w)$ value has been normalized by the $\beta(w=1/3)\equiv \beta_\text{rad}$ value. The effect becomes stronger for smaller $\sigma$ values.  
} \label{Figw}
\end{figure}

Let us now turn to the parameter $\beta$, which is of central importance for the PBH cosmology.  It is the fraction of $\rho_\text{tot}$ that collapses into black holes and it can also be interpreted as the  probability of such an   event to happen. Assuming Gaussian statistics, the black hole formation probability for a spherically symmetric region is \cite{Press:1973iz} 
\begin{equation} \label{brad}
\beta_\text{}(M)=\int_{\delta_c}d\delta\frac{1}{\sqrt{2\pi\sigma^2(M)}} e^{-\frac{\delta^2}{2\sigma^2(M)}}\,,
\end{equation}
that is approximately equal to  $\beta \sqrt{2\pi} \simeq \sigma/\delta_c e^{-\frac{\delta^2_c}{2\sigma^2}} $ for $\sigma \ll \delta_\text{max}$, where $\delta_\text{max}$ the upper value of the integration interval. 
During kination pressure is maximal and we expect the overdense regions to be spherically symmetric collapsing nearly immediately after horizon entry.
The PBH abundance has an exponential sensitivity to the variance of the perturbations $\sigma(M)$ and to the threshold value $\delta_c$.   
In the comoving gauge Ref. \cite{Harada:2013epa} finds that  $\delta_c$ has the following dependence on  $w$,
\begin{equation} \label{dc}
\delta_c \,(w)= \frac{3(1+w)}{5+3w}\sin^2 \frac{\pi \, \sqrt{w}}{1+3w}\,,
\end{equation}
 see Figure \ref{Figw}.
For the conventional radiation cosmology,
 $w=1/3$, it is $\delta_c=0.414$. In our case of kination dominated early universe   the PBH formation occurs when the density perturbation exceeds the threshold, 
 \begin{align}
 \left.\delta_{c}\right|_{w=1}=\frac38 \,.
 \end{align}
 
 Notice that the expression (\ref{dc}) is different from the  result of Carr \cite{Carr:1975qj}. In Carr's analysis an overdensity collapses if its size at the maximum expansion is smaller than the particle horizon $R_\text{hor}$ and larger than Jeans length $R_\text{J}\sim \sqrt{w} R_\text{hor}$. The former is comparable with the curvature scale and for a pure kination phase, where the sound speed is equal to one,  Jeans length and curvature scale are equal. According to Carr's condition, it is $\delta_c\simeq w$ and the pressure gradient force seems to rule out any collapse of a density perturbation during a pure kination phase. 
 However,  Carr's result has been refined in Ref. \cite{Harada:2013epa} where it was pointed out that collapse is realized if the sound crossing time over the radius is longer than the free fall time in the interval from the maximum expansion to complete collapse.  In their analysis the Jeans length has been identified with \cite{Harada:2013epa}
 \begin{align}
 R_\text{J} = a_\text{max} \sin\left( \frac{\pi \sqrt{w}}{1+3w}\right)\,,
 \end{align}
 where  $a_\text{max}$ coincides with the Hubble radius of the background Friedmann universe in the uniform density slice. 
This expression for the Jeans length implies that the threshold value of primordial black hole formation is given by Equation (\ref{dc}). It is interesting  that the analytically found $\delta_c$ value reaches its maximum for $w\sim 0.4$ and decreases as $w$ increases. The analytic $\delta_c$ decrease for large $w$ values is caused by the shortening of the dynamical time of the collapse due to the contribution of the pressure to the source of gravity.
 We note that the decrease of the threshold value for large $w$ values ($w\gtrsim 0.4$) has not been observed in numerical simulations completed for $0.01\leq w\leq0.6$ in Ref. \cite{Musco:2012au}, and the analytic expression (\ref{dc}) might underestimate the threshold value. 
 Nonetheless, what is important for us is that the threshold value for large $w$ appears to be less than the maximum allowed, $(3+3w)/(5+3w)$, and a collapse during  kination domination can be in principle realized.
 
Taking the analytic result (\ref{dc})  at face value, we see that  for $w\gtrsim 0.4$  we have a quite similar behavior to the case where the equation of state softens by taking values $w<1/3 $ as, e.g., 
during the QCD phase transition where an enhancement of the PBH formation probability is expected \cite{Byrnes:2018clq}. 
In our case the EOS becomes harder due to the stiff fluid domination and there is a $\delta_c(1/3) - \delta_c(1)=3.8\%$ decrease in the analytic value for the threshold.
   Smaller values of $w$ but still close to one  also give an enhanced PBH formation probability, $\delta_c \,(0.9)=0.385$, $ \delta_c \,(0.8) \simeq 0.395$. Thus, a smaller amplitude for the density perturbations is required during kination regime, see Figure \ref{Figw}.
We recall that the power spectrum of the comoving curvature perturbation is
\begin{align}
  {\cal P_R}=\left( \frac{5+3w}{2+2w}\right)^2 {\cal P}_{\delta}\,
\end{align}
where $\sigma^2 \sim {\cal P}_{\delta}$ is 
the variance of the density perturbations in a window of $k$.
For kination domination it is $\sigma^2 \sim (1/4)   {\cal P_R}$ and approximately
 we obtain ${\cal P_R} \sim  4 \delta^2_c/ W_0[(2 \pi \beta^2)^{-1}]$, where $W_0$ is
the principal branch of Lambert $W$ function.

\section{PBH  Evaporation and PBH Remnants}\label{S3}

The generation of PBHs via the collapse of large inhomogeneities at the time of reentry require a particular shape for the  spectrum of the primordial curvature perturbations, ${\cal P_R}(k)$. The spectrum has to be nearly flat  at scales observed at the CMB sky, $k\sim 0.05$~Mpc$^{-1}$,  with value  ${\cal P_R}\sim 2\times 10^{-9}$. On the other hand, PBHs form only if the power spectrum is amplified roughly by seven orders of magnitude at the wavenumber $k_\text{peak}$ and this amplification must not, by any means, spoil the CMB predictions. For the class of inflationary models that predict a spectral index value $n_s\sim 1-2/N_*$ and tensor-to-scalar ratio  $r\sim 12 \alpha/N_*^2$, such as $\alpha$-attractors, the Planck 2018  measured value (in the 68$\%$ confidence region),
\begin{equation}
n_s=0.964 \pm 0.004
\end{equation}
constrains significantly the position of the ${\cal P_R}$ peak on the $k$-space. 
 In Ref. \cite{Dalianis:2019asr}  it was pointed out that this class of inflationary models generate PBHs and predict 
 a spectral index value inside the 68$\%$ CL region of the Planck 2018,  
\begin{equation} \label{ns}
n_s \gtrsim 0.96
\end{equation} 
only if PBHs are ultra-light with  mass   $M_\text{PBH} \lesssim10^5$ grams \cite{Dalianis:2019asr}.
This rough bound comes from (\ref{ns}) and the relation $\Delta N \sim 2/(1-n_s)$, where $\Delta N$ are the efolds of inflation that take place in between the moments of horizon exit of the  scales $k=0.05$ Mpc$^{-1}$ and $k_\text{peak}$. 
$M_\text{PBH}$ admits larger values for $n_s>0.955$ which is the lower bound of the 95$\%$ CL region. 
This issue has been also discussed recently in \cite{Iacconi:2021ltm}, where they found agreement with the CMB data for  $M_\text{PBH}\lesssim10^8$g in the framework of $\alpha$-attractors.

The fact that ultra-light  PBHs seem to be favored by the CMB measurements of the spectral index $n_s$ and the running of the spectral index $\alpha_s$ prompt us to investigate  inflationary models that generate  mini PBHs. 
The striking characteristic of the mini PBH scenario is first, the Hawking radiation emitted by the evaporating PBHs and second, the remnants that might be left behind at the end of the evaporation process.

\subsection{Evaporation}
Hawking has shown  that black holes radiate thermally with a temperature \cite{Hawking:1975vcx, Hawking:1974rv}
\begin{align} \label{HawkT}
T_\text{BH}&=\frac{\hbar c^3}{8\pi G M_\text{BH} k_B} \nonumber \\
&=\frac{M^2_\text{Pl}}{M_\text{BH}}=1.05 \times  10^4 \left( \frac{M_\text{BH}}{10^{9}\text{g}}\right)^{-1} \text{GeV}\,.
\end{align}
assuming the Schwarzschild solution, that is a black hole without charge or angular momentum. Black holes  are expected to evaporate on a time scale $t_\text{evap} \sim G^2 M_\text{BH}^3/(\hbar c^4)$.
  In the second line of Equation (\ref{HawkT}) we took $k_B=c=\hbar=1$ and  we will use this convention in the following.

\begin{figure}[!htbp]

\includegraphics[width = 0.68\textwidth]{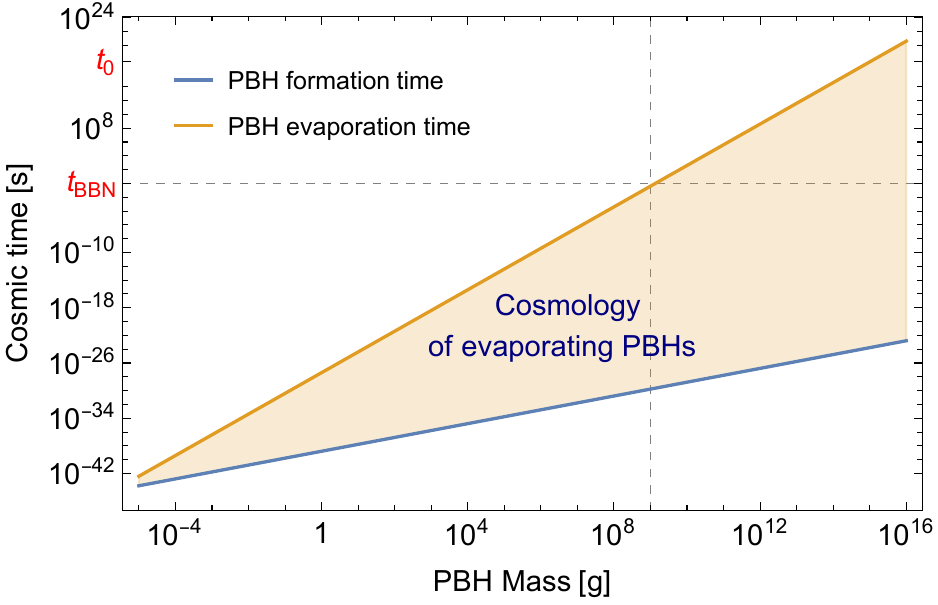}
\caption{The mass range and timescales of 
 evaporating  PBH cosmology. 
The lifetime of the universe and the cosmic time of BBN are highlighted. The former subdivides the evaporating PBHs from the dark matter PBHs;    the latter 
defines a threshold upper  mass scale, $M_\text{PBH}\sim 10^9$g,  where  PBHs evaporation does not upset the hot big bang observables.} \label{FigPBHcosmo}
\end{figure}

The mass-loss rate of an evaporating black hole is 
\begin{align} \label{dMpbh}
\frac{d M_\text{BH}}{dt}=-7.6 \times 10^{6} \,  {g_\text{H}(T_\text{BH})} \left( \frac{M_\text{BH}}{10^{9}\text{g}}\right)^{-2} \,\text{g}\,\text{s}^{-1}
\end{align}
where ${g_\text{H}(T_\text{BH})} $ is a spin-weighted number of degrees of freedom of emitted particle species and it takes the value 108 for the Standard Model particles for hot enough black holes, $T_\text{BH}>100$ GeV or $M_\text{BH}<10^{11}$ g; in the limit of  cool black holes it is ${g_\text{H}(T_\text{BH})} \sim 7$.
 Integrating the mass-loss rate over time, the time dependent mass of the BH,
 
 \begin{align}
 M_\text{BH}(t) \simeq  M_\text{BH}(t_\text{f})\left(1-\frac{t}{t_\text{evap}} \right)^{1/3},
 \end{align}
for $t_\text{f}\ll t_\text{evap}$, and the BH lifetime, 
\begin{align} \label{tevap}
t_\text{evap}(M_\text{BH}) \simeq 0.41 \left(\frac{g_\text{H}(T_\text{BH})}{108} \right)^{-1} \left( \frac{M_\text{BH}}{10^{9} \text{g}} \right)^3 \text{s},
\end{align}
are found \cite{Carr:2009jm}.
For $t_\text{evap}$ equal to the present age of the universe, $13.8$ billion years, one finds a critical black hole mass $M_\text{BH, cr}\simeq 10^{15} (g_\text{H}(T_\text{BH})/108)^{1/3}\sim 5 \times 10^{14}$ g. Apparently, light BH with mass $M_\text{BH}\ll M_\text{BH, cr}$ are irrelevant for late time cosmology but might play a crucial role in the early cosmic evolution, see Figure \ref{FigPBHcosmo}.

PBHs with mass $M_\text{PBH} \sim M_\text{BH, cr}$ are a source of gamma rays today and can constitute only a small fraction of the dark matter in the galaxies. Lighter PBHs generate an entropy at the evaporation that might alter big bang observables such as BBN. 
In particular, for lifetimes $\tau=10^{2}$--$10^{7}$ s, which correspond to $M_\text{PBH}=10^{10}$--$10^{12}$ g, hadrodissociation processes become important and the debris deuterons and nonthermally produced $^6$Li constrain $\beta(M_{\text{PBH}})$;  for lifetimes
$\tau= 10^7 $--$10^{12}$  s, which correspond to \mbox{$M_\text{PBH}=10^{12}$--$10^{13}$~g}, 
 photodissociation processes 
  overproduce  $^3$He and D  and put strong constraints on $\beta(M_\text{PBH})$~\cite{Page:1976wx, MacGibbon:1991vc, Carr:1998fw,  Carr:2016hva}.
In addition, the heat produced by PBHs evaporation after the era of recombination 
damps small-scale CMB anisotropies
and 
in the  mass range $ 2.5 \times 10^{13}~\text{g} \lesssim M_\text{PBH} \lesssim 2.4 \times 10^{14}~\text{g} $   only a very suppressed PBH abundance is allowed~\cite{Chen:2003gz, Zhang:2007zzh, Carr:2009jm}.
The conclusion is that PBHs with mass in the range $10^9$--$10^{17}$ g are significantly constrained~\cite{Carr:2020gox}.  In the smaller mass limit the tight constraints on PBHs abundance are raised. 

Finally, let us add  some interesting remarks.  First,  if the temperature of a PBH is initially smaller than the background cosmic temperature, accretion effects should be taken into account, but this practically should not modify PBH lifetime.
Accretion decreases the temperature of a PBH but this decrease is negligible  for small enough PBH mass, whereas the cosmic temperature falls  fast due to the expansion. 
Second, the Hawking temperature  $T_\text{BH}$ is expected to reach a maximal value at some point and  afterwards decrease. In this last and most uncertain stage of the PBH evaporation the rate $dM/dT_\text{BH}$ should turn into~positive.

\subsection{Remnants from the Evaporation of PBHs }

Apparently the evaporation process of  BHs is uncertain at masses of the order of the Planck scale. 
The semiclassically derived Hawking temperature $T_\text{BH}$ diverges for vanishing $M_\text{BH}$ and there is a logical guess that Hawking radiation halts somewhere near the Planck scale, leaving behind a stable black hole remnant\footnote{In the literature they  are also called BH relics; however, we avoid this term here since massive PBHs that have not evaporated by now ought to be called, also, big bang relics. } with mass $M_\text{rem}$. 
The  PBH remnant mass 
can be written in terms of the Planck mass
\begin{align}
M_\text{rem}={\cal K} \, m_\text{Pl} ,
\end{align}
where ${\cal K}$ is a factor that parameterizes our ignorance of the physics that operates at the relevant energy scales.
Energy conservation \cite{Torres:2013kda}, extra spatial dimensions \cite{ArkaniHamed:1998rs, Suranyi:2010yt},  higher order corrections to the action of general relativity \cite{Barrow:1992hq}, the information loss paradox \cite{Chen:2014jwq}  might prevent complete evaporation.

Different theories predict stable black hole relics of different mass. 
${\cal K}$ may be of order one, with remnant masses characterized by  the fundamental scale of gravity, $m_\text{Pl}=G^{-2}$,  but other values for $\cal K$ are also admitted. It can be ${\cal K}>1$ if BH have quantum hair \cite{Coleman:1991ku},
or ${\cal K}<1$ in  treatments where a generalized uncertainty principle is applied \cite{Adler:2001vs}. 
In our analysis we let $\cal K$ be a free parameter and we remain agnostic about the fundamental physics that might prevent black holes from complete evaporation.

Equation (\ref{dMpbh}) written in the compact form $\dot{M}_\text{BH}=-A \,M^{-2}_\text{BH}$ yields the time dependence of the black hole mass
\begin{align}
M^3_\text{BH}(t)\approx 3A\,(t_\text{evap}-t) +M^3_\text{rem}
\end{align} 
for $t\leq  t_\text{evap}$ and $A$ a dimensionful parameter dependent on $T_\text{BH}$, 
$A=\pi {\cal G} g_H(T_\text{BH}) M^4_\text{Pl}/480$, where ${\cal G}$ is the gray-body factor.
The decay of the BH mass implies that we can define a decay rate, $\Gamma_\text{BH}$, according to the formula $\dot{\rho}\equiv -\Gamma_\text{BH} \,\rho$. We attain $\Gamma_\text{BH} =  -\dot{M}_\text{BH}/M_\text{BH}$ and, contrary to the conventional perturbative particle decays,  here it is the black hole  mass that decays not the number density.
The time dependent decay rate is 
\begin{align} \label{GammaBH}
\Gamma_\text{BH}(t)=\frac{A}{M^3_\text{BH}(t)} \, = \,\frac{1}{3(t_\text{evap}-t)+c_{\cal K} t_\text{Pl} }\,,
\end{align}
where $c_{\cal K}={\cal K}^3 480 \sqrt{8\pi}/(\pi {
\cal G} g_H(T_\text{BH}))$. As $t \rightarrow t_\text{evap}$ the final explosive phase of black hole evaporation takes place.

At the moment right after the PBH evaporation the energy  of a PBH remnant is 
$E_\text{rem}(t^+_\text{evap}) = \sqrt{\vec{p}^{\,2}_\text{rem}+M^2_\text{rem}}$
where $\vec{p}_\text{rem}={M_\text{rem} \vec{v}_\text{rem}}/{\sqrt{1-\vec{v}^2_\text{rem}}}$
 is the momentum obtained by the remnant after the completion of the evaporation,
with $\vec{v}_\text{rem}$  the recoil velocity of the remnant. The order of magnitude of the recoil velocity can be determined only after particular assumptions about the final stage of BH evaporation are made. 

As discussed above, it is expected -but not firmly confirmed- that the result for the Hawking temperature, Equation (\ref{HawkT}), stops being valid in the vicinity of Planck densities.
Let us assume that there is a maximal temperature $T_\text{max}$ for the black hole during the course of evaporation and afterwards a cooling down occurs during the  emission process. 
 A maximum temperature implies the presence of a black hole remnant of mass $M_\text{rem}$, that might  or might not be a horizonless object. 
Accordingly, the surface temperature of the remnant, $T_\text{rem}$, might be zero or in equilibrium with the background cosmic temperature. In both cases, the temperature of the final state is well below the Planck energy scale, whereas it is plausible to assume that $T_\text{max}$ is not far from  the Planck scale. 
Hence we take $T_\text{max}-T_\text{rem} \sim T_\text{max} \lesssim {\cal O} (M_\text{Pl})$. 
From the moment where  $T_\text{BH}=T_\text{max}$ until the moment of the remnant formation a number of  $N_f$ quanta has been emitted. 
The typical energy of these quanta is $E\sim T_\text{max} \sim \Delta M/ N_f$ where $\Delta M=M_\text{BH}(T_\text{max})-M_\text{rem}$ is the mass radiated away in the final stage of the evaporation. The number $N_f$ of quanta depends on the model assumed for the the description of the final stage. 
 Ref. \cite{Kovacik:2021qms}  discusses these issues and gives a number $N_f\leq {\cal O}(10^2)$.
The momentum of each quantum emitted is about $\Delta M /N_f$ and the evaporating black hole performs a random walk in the momentum space leaving behind a remnant with average momentum \cite{Kovacik:2021qms} $p_\text{rem}(t^+_\text{evap})\sim {\Delta M}/{\sqrt{N_f}}$
 and recoil velocity, 
\begin{align}\label{vrem}
v_\text{rem}(t_\text{evap})\sim \frac{\Delta M}{M_\text{rem}}\left(N_f+\left(\frac{\Delta M}{M_\text{rem}}\right)^2\right)^{-1/2}.
\end{align}

\subsubsection*{Remnants  CDM Scenario}

The remnants, though initially close to relativistic, are  cold dark matter today.  
The momentum of remnants redshifts as $a^{-1}$, and being kinetically decoupled, they free-stream a distance
\begin{align}
k^{-1}_\text{fs} & = \int_{t_\text{evap}}^{t_0}  
\frac{v_\text{rem}(t_\text{evap})}{a(t)} dt \nonumber \\
&
= \int_{a_\text{evap}}^1 \frac{p _\text{rem}}{E_\text{rem} } \frac{1}{a^2 H} da\nonumber \\
&\sim 5.6\, \text{Mpc} \left( \frac{v_\text{rem}(t_\text{evap})}{0.1} \right)\left( \frac{a_\text{evap}}{a_\text{eq}} \log\left(\frac{4a_\text{eq}}{a_\text{evap}}\right)\right)\,,
\end{align}
where
subscript ``0” denotes present values and ``eq” equality era.
It has to be $k_\text{fs}^{-1}<0.1\,\text{Mpc}$ in order that the smallest structures observed in the matter power spectrum are not washed out by the free stream of remnants.
This constraint translates into an upper bound on $t_\text{evap}$, which, in any case, precedes the BBN epoch. 
As noted in \cite{Lehmann:2021ijf},  the remnant dark matter scenario is a CDM scenario in agreement with the cosmological observations.

Next we overview the cosmological evolution of mini PBHs  and their remnants.

\section{Evolution of Mini PBHs and Their Remnants during a Kination Era}\label{S4}

In this section, we will discuss elements of the cosmological evolution of mini PBHs and their remnants with the background energy density dominated by a stiff fluid. Initially, we will examine the exact system of equations that describes the evolution of the different cosmic fluids. Afterwards,
we will pursue an approximate analytic description that will show the characteristic mass, time and temperature scales of our scenario and the parameter space that a viable cosmology is realized.

\subsection{Energy Densities}

\begin{figure}[!htbp]
  \includegraphics[width=0.496\linewidth]{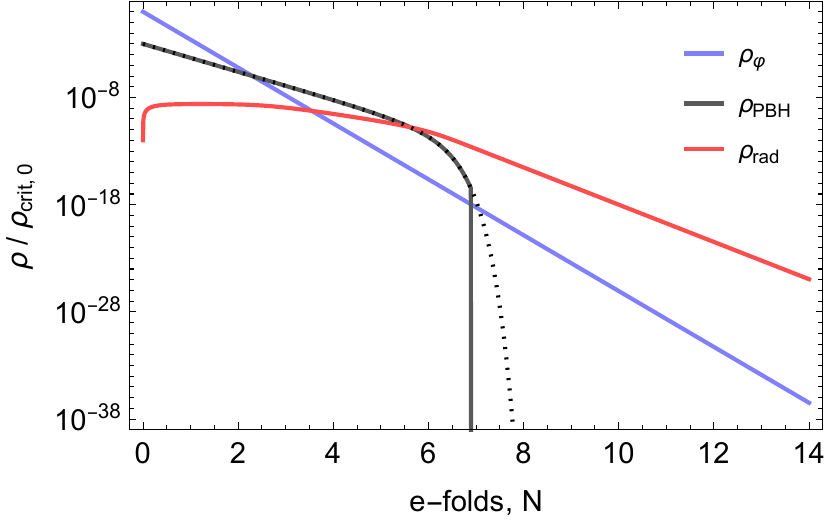}
  \includegraphics[width=0.49\linewidth]{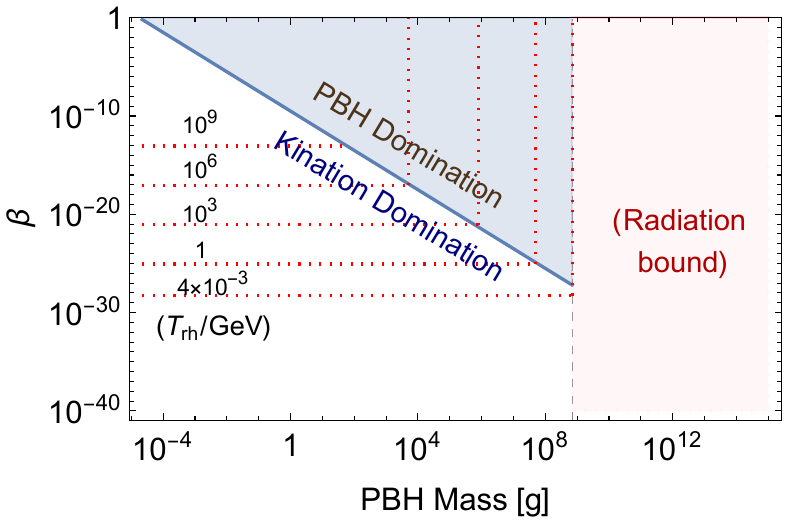}
\caption{In the \textbf{left panel} the evolution of the energy densities of a runaway scalar field,  PBHs  and radiation, $\rho_\varphi$, $\rho_\text{PBH}$ and $\rho_\text{rad}$, respectively, normalized by the initial critical energy density $\rho_\text{crit, 0}$, are shown for a transit PBH domination phase.
The runaway scalar field is modeled as a stiff fluid, the PBHs initially evolve as a pressureless fluid  but they promptly evaporate producing radiation.  
The truncation of the PBH evolution is dictated by Equation (\ref{GammaBH}); for comparison the effect of a  constant in time  decay rate is also depicted.
The radiation initially evolves as $N$ and after the completion of the evaporation as $e^{-4N}$.
 In the \textbf{right panel} we depict the $\beta$ values  that a PBH domination phase is realized/avoided for a background  dominated  by stiff fluid.
We also contour plot five values of the reheating temperature due to PBH evaporation.
} \label{Fig02bounds}
\end{figure}

We assume a universe initially dominated by the kinetic energy of the inflaton scalar field $\varphi$. Its energy density has a maximal redshift $a^{-6}$.
During the kination phase the production of PBHs takes place with $\beta\ll 1$, that is the initial amount of PBHs is tiny compared to the bulk energy density. The PBHs evolve  as pressureless matter $a^{-3}$ until the cosmic time  becomes comparable to the evaporation time, $t_\text{evap}$.
 Assuming that the evaporation of PBHs results in relativistic particles mostly, leaving also behind a remnant mass,  we have a universe with energy density partitioned among three fluids: scalar field, evaporating PBHs and radiation in the first phase, and  scalar field,  radiation and  PBH remnants in the second phase. This dynamical system is described by the Friedmann together with the continuity equations, 
 
 \begin{align}
 \frac{d\rho_\varphi}{dN} &\, = \, -6\rho_\varphi \,,\\
 \frac{d\rho_\text{PBH}}{dN} &\, =\,-3\rho_\text{PBH} -\frac{\Gamma_\text{BH}}{H} \rho_\text{PBH} \,,\\
 \frac{d\rho_\text{rad}}{dN} &\, =\,-4\rho_\text{rad} +\frac{\Gamma_\text{BH}}
 {H}\rho_\text{PBH} \, \\
 \frac{dH}{dN}& \, = \,-\frac{1}{2H M^2_\text{Pl}}\left( 2\rho_\varphi+\rho_\text{PBH}+\frac{4}{3} \rho_\text{rad} \right)\,, \label{systH}
 \end{align}
 where  $dN=d(\ln a)=Hdt$ is the differential of the e-fold number.
 Figure \ref{Fig02bounds} depicts the evolution of this system of fluids with a transit PBH domination phase and for arbitrary initial conditions. For $t\ll t_\text{evap}$  the PBH energy density scales such as $e^{-3N}$ and that of radiation increases as $N$. 
   The scalar field rolls down its steep potential without decaying, but redshifting as $e^{-6N}$. 
The produced radiation will dominate the energy density reheating finally the universe.    
   In the late universe the dark energy of the runaway scalar field $\varphi$ might be the cause of the accelerated expansion observed today.

 After the PBH evaporation, the remnant of each PBH  is left behind. Their energy density evolves as 
 \begin{align}
  \frac{d\rho_\text{rem}}{dN}  \,\approx\, -3 \rho_\text{rem}\,, \quad \text{for} \,\,\,  t>t_\text{evap} 
 \end{align}
 
 It is understood that in Equation (\ref{systH}), $\rho_\text{PBH}$ is replaced  by the remnant energy density. 
The PBH remnants will be dark matter constituents of  the galaxies today.

The exact evolution can be traced by solving the above system of equations. 
In the following, for clarity and simplicity, we will make the approximation of instantaneous evaporation for  PBHs. This way, analytic expressions for the energy densities and the constraints, as well as threshold conditions, can be obtained.

\subsection{Reheating the Universe via PBH Evaporation}

Runaway inflation scenarios in their simplest version cannot reheat the universe.
A model-independent source of radiation comes from the de Sitter horizon that is characterized by a Hawking temperature. If radiation domination is realized by this process, gravitational reheating is achieved \cite{Ford:1986sy, Chun:2009yu},  but it is rather inefficient. A completion  with  specially designed interactions is necessary in order  for the thermalized early universe to be realized, see Appendix \ref{A1}.
 In our scenario, the Hawking radiation from mini PBHs automatically reheats the universe. The special characteristic is the need of a ``feature”, such as an inflection point or a step, which enhances the curvature power spectrum. 

Let us start initially assuming  a general EOS $w$ for the background.
In the approximation of instantaneous evaporation, the moment right before evaporation,  $t^-_\text{evap}$, the energy density of the PBHs over that of the scalar field   is 
\begin{align} \label{thresh0}
\frac{\rho_\text{PBH}(t^-_\text{evap})}{\rho_\varphi(t^-_\text{evap})}
& \approx 
\gamma \, \beta\, \tilde{g}\,\left(  \frac{ M_\text{hor}(t^-_\text{evap})}{M_\text{PBH}/\gamma}  \right)^\frac{2w}{1+w} \,,
\end{align}
 where we considered  $a(t)\propto t^{\frac{2}{3(1+w)}}$ and 
 we  plugged in the horizon mass.
 The coefficient
 $\tilde{g}=\tilde{g}(g_*, t^-_\text{evap})$ is equal to one unless 
the universe is radiation dominated; in the latter case it is $\rho_\text{rad}\propto g_* T^4$ and $T\propto g_s^{-1/3}a^{-1}$ 
where $g_s$  are the entropic degrees of freedom and $g_*$ the  thermal ones. 
Substituting the mass inside the curvature radius at the evaporation moment of the  PBH 
$M_\text{hor}(t_\text{evap})=3 M_\text{PBH}^3(1+w)/4m^2_\text{Pl}$, a threshold  $\beta(M_\text{PBH})$ value is found
that specifies whether or not the universe has  become PBH dominated. 

Let us assume that the formation of black holes and the subsequent evaporation takes place during an early  cosmic era of stiff fluid domination,   called kination dominated phase (KIN).
In our context it is the inflaton field $\varphi$ that rolls down a runaway potential  that gives rise to kination phase. 
 According to Equation (\ref{thresh}) and for $w=1$ the energy density of the PBHs at the moment right before evaporation is 
\begin{align} \label{KinEvap}
\frac{ \rho_\text{PBH}(t^-_\text{evap})}{\rho_\varphi(t^-_\text{evap})} \simeq \frac{3}{2}\gamma^{2}_\text{} \beta \frac{M^2_\text{PBH}}{ m^{2}_\text{Pl}} \,.
\end{align}  

Neglecting the PBH mass loss due to the evaporation process, the assumption of a kination phase is valid roughly  for $\gamma^{2}_\text{} \beta  < m^{2}_\text{Pl}/M^2_\text{PBH}$,  otherwise a PBH dominated universe is realized.  Plugging in benchmark values, the PBH domination is avoided for $\beta<\beta_\text{thresh}$~where,
\begin{align}\label{thresh}
\beta_\text{thresh} \, =  \,0.3 \times 10^{-19} \gamma^{-2} \left( \frac{M_\text{PBH}}{10^{5} \text{g}}\right)^{-2}\,.
\end{align}

In Figure \ref{Fig02bounds}, we depict the bound above and highlight the parameter space that a kination or a PBH domination phase is realized for PBHs with mass $M_\text{PBH}<10^9$ g.

\subsubsection{Reheating without PBH Domination}

After PBH evaporation the background energy density is partitioned between the runaway inflaton, $\rho_\varphi$, the entropy produced by the PBH evaporation, $\rho_\text{rad}$ and remnants $\rho_\text{rem}$. 
Radiation  is about $M_\text{PBH}/M_\text{rem}$ times larger than  $\rho_\text{rem}(t_\text{evap})$. Assuming  fast thermalization of the evaporation products, the radiation redshifts like $\rho_\text{rad} \propto g_* g_s^{-4/3} a^{-4}$ and sooner or later it will dominate  the runaway scalar field  that redshifts as $\rho_\varphi \propto a^{-6}$. 
At  some  moment the radiation dominates the energy budget of the universe and  when $\rho_\varphi (t)=\rho_\text{rad}(t)$  we define the reheating 
 moment $t=t_\text{rh}$. 
Equivalently, a reheating temperature of the universe is defined that reads  \cite{Dalianis:2019asr}
\begin{align}
 \label{Tkination}
T_\text{rh} \approx 10^{-2} \, \text{GeV} \left( \frac{\beta}{10^{-28}} \right)^{3/4} \gamma^{3/2}_\text{}  g_*^{-1/2} \,.
\end{align}

It is notable that the reheating temperature depends only on the $\beta$ value.

\subsubsection{Reheating after PBH Domination}

Let us now consider the case $\beta > \beta_\text{thresh}$ where PBHs dominate the energy density of the universe before evaporation.  
At the time $t^{-}_\text{evap}$ we can approximate $\rho_\text{PBH}\approx 3 H^2 M^2_\text{Pl}$ and at the moment of evaporation nearly the entire $\rho_\text{PBH} $ transforms into radiation with density $\pi^2 g_* T_\text{rh}^4/30$ and $H=2/(3t)$. The reheating temperature  of a PBH dominated universe is
\begin{align} \label{TrhPBHdom}
T_\text{rh}  \approx 10^3 \text{GeV} \left( \frac{10^5 \text{g}}{M_\text{PBH}}\right)^{3/2} \left(\frac{106.75}{g_*(T_\text{rh})}\right)^{1/4} \,.
\end{align}

Comparing with Equation (\ref{Tkination}), we see that $T_\text{rh}\propto M_\text{PBH}^{-3/2}$ and does not depend on $\beta$.
 In the left panel of  Figure \ref{Fig02bounds} some benchmark $T_\text{rh}$ values are labeled on the $(M_\text{PBH}, \beta)$ parameter space  both for kination and PBH domination phase.

\subsection{PBH Remnants  Abundance }

The moment right after the evaporation, which we label $t^+_\text{evap}$, the energy density of the PBHs has either vanished or a remnant is left behind. In the latter case the evaporation halts  somewhere close (or not) to the Planck scale
and the energy density contained in the form of  remnants is 
 $(M_\text{rem}/ M_\text{PBH}) \rho_\text{PBH}(t^-_\text{evap})$, plus their kinetic energy.
 The factor    $(M_\text{rem}/ M_\text{PBH})$ is much smaller than one and nearly the entire energy density of the initial population of PBHs turns into radiation apart from a tiny amount. 
The PBH remnants have a number density,
\begin{align}
n_\text{rem}(t^+_\text{evap})=\frac{\rho_\text{tot}}{M_\text{PBH}}\left(1+\frac{m^2_\text{Pl}}{\frac32 \gamma^2 \beta M^2_\text{PBH}}\right)^{-1}
\end{align}
where $\rho_\text{tot}=\rho_\varphi+\rho_\text{rad}+\rho_\text{rem}$ is the sum of energy densities for the runaway inflaton, radiation and remnants.

 \begin{figure} [!htbp]
  \includegraphics[width=.49 \linewidth]{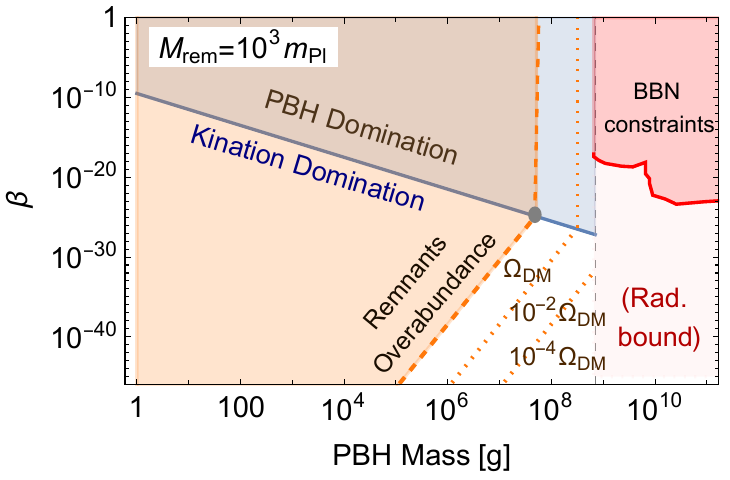}
  \includegraphics[width=.49\linewidth]{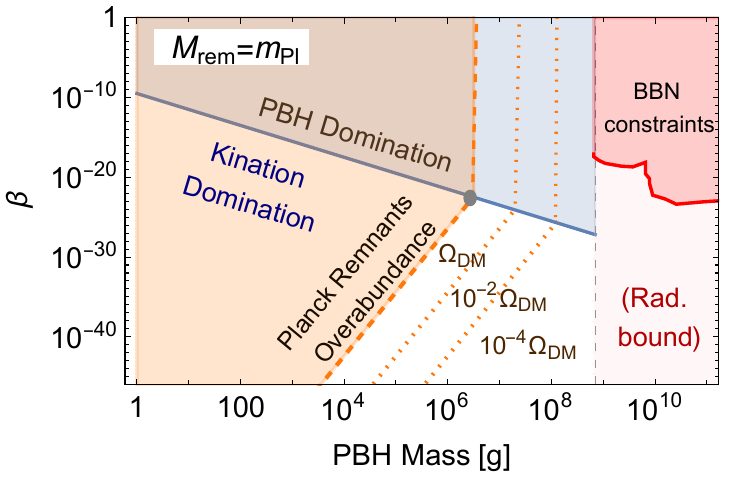}\\
  \includegraphics[width=.49 \linewidth]{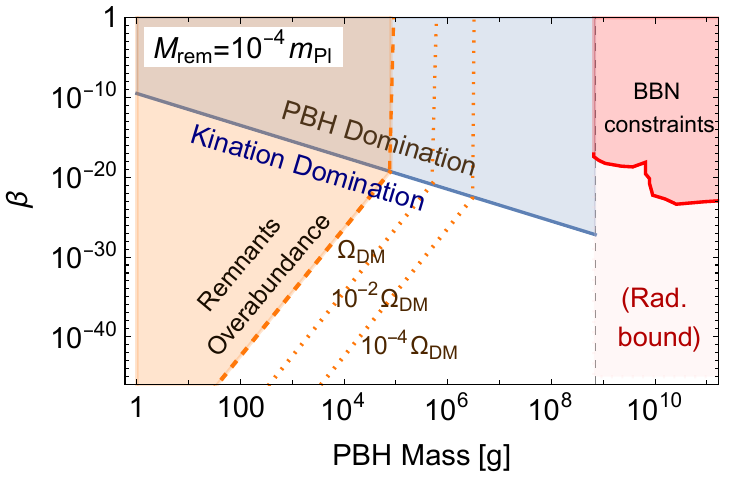}
  \includegraphics[width=.49\linewidth]{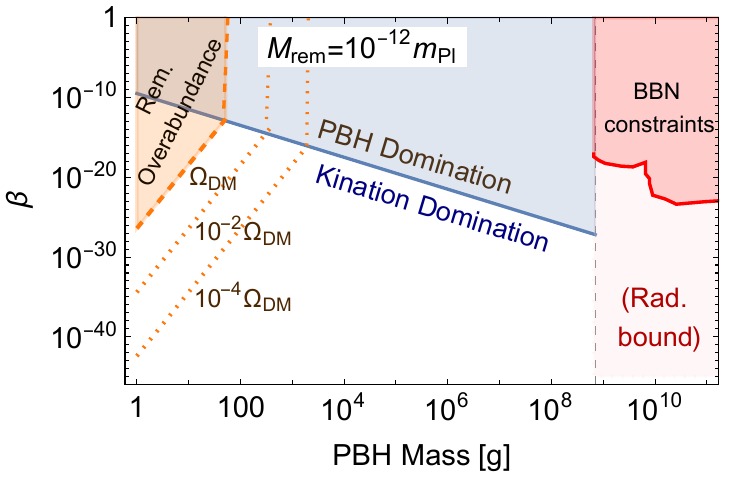}
\caption{In each panel  the white and colored regions depict the parameter space  that  gives viable cosmology  without overabundance of remnants. 
The four panels correspond to four different remnant masses $M_\text{rem}/m_\text{Pl}=
10^3, 1, 10^{-4}, 10^{-12} $, respectively. The  orange dotted lines give the $(M_\text{PBH}, \beta)$ values that realize the PBH remnant dark matter scenario and on the dashed line the $\Omega_\text{DM}$ bound is saturated.
} \label{Figremnants}
\end{figure}

\subsubsection{Remnants Abundance without PBH Domination }
Let us estimate the remnants abundance  ignoring any interaction between the different fluids and  approximating again the redshift of each fluid with the expression $\rho \propto a^{-3(1+w)}$ for constant $w$.
Until the moment  $t_\text{rh}$ the energy density of the PBH remnants increases relatively to the runaway scalar field  as $\rho_\text{rem}/\rho_\varphi\propto a^3$ and afterwards, that radiation dominates, it increases as $\rho_\text{rem}/\rho_\text{rad}\propto T^{-1}$.  
It is
\begin{align} \label{frelS1}
\frac{\Omega_\text{rem}}{\Omega_\text{DM}}=  \tilde{c}_\text{S} \frac32 \,  \gamma^{2}_\text{} \, \beta \, 
\frac{M_\text{rem} M_\text{PBH}}{m^2_\text{Pl}}
\left(\frac{a(t_\text{rh})}{a(t_\text{evap})} \right)^3 \left(\frac{M_\text{eq}}{M_\text{rh}}\right)^{1/2}
\end{align}
where $\tilde{c}_\text{S}=2^{1/4} \left({g(T_\text{rh})}/{g(T_\text{eq})} \right)^{-1/4} \Omega_\text{Matter}/\Omega_\text{DM}$
and  $M_\text{eq}\sim 6\times 10^{50}$ g is the horizon mass at the moment of radiation-matter equality. 
 For times $t<t_\text{rh}$ the Hubble radius mass increases like $M_\text{hor} \propto a^3$ and given that $M_\text{hor}(t_\text{evap})=(3/2)M_{\text{PBH}}^3/m^2_\text{Pl}$ the above  expression can be  simplified,
\begin{align} \label{fremKIN}
\frac{\Omega_\text{rem}}{\Omega_\text{DM}}
\approx 10^6 \, \sqrt{\gamma_\text{}}  \, \, \frac{M_\text{rem}}{m_\text{Pl}}\, \left(\frac{\beta}{10^{-10}}\right)^{1/4} \, 
\left( \frac{M_\text{PBH}}{10^5 \text{g}} \right)^{-2 }\,.
\end{align}

For the benchmark values $\beta\sim 10^{-10}$ and $M_\text{PBH}\sim 10^5$ g the remnants are found overabundant, except for  masses much less than the Planck mass. 
In Figure \ref{Figremnants}, we depict the allowed parameter space for the kination scenario with reheating via PBH evaporation when a remnant is left behind.  
Since $\Omega_\text{rem}/\Omega_\text{DM}\leq 1$, a large part of the $(M_\text{PBH}, \beta)$ space is ruled out because remnants
overclose the universe.
Smaller $\beta$ values are in accordance with the observed dark matter  values  for $M_\text{rem}\sim m_\text{Pl}$. However, such $\beta$ values are rather small for a successful reheating. We will further comment on the allowed $\beta$ values in the~following.

\subsubsection{Remnants Abundance with PBH Domination }
 
If the fraction of the universe energy density that collapses into PBHs is above $\beta_\text{thresh}$ then a PBH domination phase is realized. 
 At the moment of PBH evaporation the remnants contain a small amount of the energy density,
\begin{align}
\left. \rho_\text{rem}\right|_{t^{+}_\text{evap}} \approx \rho_\text{rad}(T_\text{rh}) \frac{M_\text{rem}}{M_\text{PBH}},
\end{align}
where $T_\text{rh}$ is given by Equation (\ref{TrhPBHdom}).
 Taking into account that $\rho_\text{rad} \propto g_* g_s^{-4/3} a^{-4}$, $\rho_\text{rem}\propto a^{-3}$  
 and 
 Equation (\ref{tevap}) that gives the time of evaporation,
 we find the remnants abundance~today,
 \begin{align} \label{fremMD1}
 \frac{\Omega_\text{rem}}{\Omega_\text{DM}}
\approx  \frac{M_\text{rem} \, m_\text{Pl} \, \sqrt{M_\text{eq}}}{M_\text{PBH}^{5/2}} \left( \frac{g_*(T_\text{eq})}{g_*(T_\text{rh)}}\right)^{1/4}\,,
 \end{align}
where we took the entropic and thermal degrees of freedom equal. After normalizing with benchmark values we rewrite, 
\begin{align} \label{fremMD}
\frac{\Omega_\text{rem}}{\Omega_\text{DM}}
\,\sim \, 10^3  \,  \frac{M_\text{rem}}{m_\text{Pl}}\,
\left( \frac{M_\text{PBH}}{10^5 \text{g}} \right)^{-5/2 } \left(\frac{106.75}{g_*(T_\text{rh})} \right)^{1/4}\,.
\end{align}

We note that the  expression (\ref{fremMD1}) can also be derived from Equation (\ref{fremKIN}) after plugging in the expression for $\beta_\text{thresh}$.

The PBH remnants  constitute a significant part of the cosmic dark matter only if 
$M_\text{BBH}\sim 10^6(M_\text{rem}/m_\text{Pl})^{2/5}$. Since it must be $M_\text{PBH}<10^9$ g, the remnants cannot have a mass larger than $10^7 m_\text{Pl}$. Otherwise, the remnants have a small or negligible contribution  to the total dark matter abundance.

\subsection{Constraints from Inflation}

\subsubsection{Minimum Masses}

Right after the end of inflation the Hubble horizon mass has a minimum value $4\pi \rho_\text{end} H^{-3}_\text{end}/3$ and steadily grows. 
We consider PBHs that form from the collapse of strong  overdensities  that reenter the horizon (different scenarios have been also considered, see, e.g., \cite{Nakama:2018lwy, Martin:2019nuw}). Hence, PBHs cannot have mass smaller than the value,
\begin{align} \label{minMpbh}
M_\text{PBH} \, \gtrsim \gamma \, \frac{4\pi M^2_\text{Pl}}{H_\text{end}} \gtrsim  \, 2 \, \gamma  \sqrt{\frac{0.036}{r_*}}\, \text{grams}\,,
\end{align}
where $r_*$ is the tensor-to-scalar ratio at CMB scales and, in the right hand side  we used that $H_\text{end}<H_*\approx (\pi^2 A_{{\cal R}*} r_*/2)^{1/2} M_\text{Pl}$. 
 We also normalized $r_*$ with the most recent bound from BICEP collaboration \cite{BICEP:2021xfz}. 
Plugging in the minimum PBH mass in Equation (\ref{fremMD}) (or in Equation (\ref{fremKIN}) and assuming maximal $\beta$ value so as to maximize $\Omega_\text{rem}$)  we find a limiting minimum mass for the remnants,
\begin{equation} \label{MinMrem}
M_\text{rem,min} \sim 10^4 \, \gamma^{5/2}  \left( \frac{0.036}{r_*}\right)^{5/4}\, \text{GeV} \,.
\end{equation}

The implication of this  cosmological bound is that  if a particle/object  with mass  less than $M_\text{rem,min}\sim 10^4$ GeV is found, then this  particle/object is unlikely to  be a remnant from PBH evaporation.

Recalling that  the PBH mass has to be
$M_\text{PBH}<10^9$ g,
 we conclude that, together with the lower bound (\ref{MinMrem}), 
 the remnants must have a mass in the range
\begin{align}
 10^{-15} \gamma^{5/2}\left(\frac{r_*}{10^{-2}} \right)^{-5/4} \lesssim \, \frac{M_\text{rem}}{m_\text{Pl}} \, \lesssim   10^{7} 
\end{align}
in order to have a non-negligible cosmic abundance.

\subsubsection{Inflaton  Residual Energy Density}

BBN is a process with distinct observables that are sensitive to modifications of the background dynamics. Energy contributions beyond radiation  must be sufficiently suppressed to prevent a too fast expansion  during BBN, 
A kination regime has two important implications.  First, $\rho_\varphi$ is non zero during BBN and second, the power spectrum of primordial GWs is blue shifted.
 Parametrizing the extra energy components  by an equivalent number of additional\footnote{There are examples of physical mechanisms  \cite{Ichiki:2002eh, Apostolopoulos:2005at} that  decrease the effective neutrino number.} neutrinos\footnote{The effective number of neutrino species present in the thermal bath after $e^+ e^{-}$ annihilation is often labeled as $N_\text{eff}$, while before $e^+ e^{-}$ annihilation as $N_\nu$. Here for simplicity we label the additional neutrino number as $\Delta N_{\nu,\text{eff}}$ either before or after $e^+ e^{-}$ annihilation.} 
\begin{align}
\rho_\text{rad}+\rho_\varphi+\rho_\text{GW}=\frac{\pi^2}{30}\left[2+\frac{7}{8}(4+2(N_\nu +\Delta N_{\nu,\text{eff}}))\right] T^4
\end{align} 
  the expansion rate  has to be less than \cite{Simha:2008zj}
\begin{align} \label{Hbbn}
\left. \left( \frac{H}{H_\text{rad}}\right)^2 \right|_{T=T_\text{BBN}} \leq 1+\frac{7}{43}\Delta N_{\nu,\text{eff}}\simeq 1+3.8 \%  \left( \frac{\Delta N_{\nu,\text{eff}}}{0.234} \right)
\end{align}
where $H_\text{rad}$ is the Hubble parameter for a universe filled only with the Standard Model radiation fluid, which 
 is made of photons, electrons, neutrinos and their antiparticles. 
In our scenario, part of the total energy density $3H^2 M^2_\text{Pl}$ during BBN is reserved by the   inflaton field and
 Equation (\ref{Hbbn}) is equivalent to  an upper bound on $\rho_\varphi/\rho_\text{rad}$ at BBN epoch.  For a field $\varphi$ that fast rolls, $\rho_\varphi=\dot{\varphi}^2/2$ a bound on the reheating temperature 
 $T_\text{rh}\gtrsim {\cal O} (10)$ MeV has been mentioned \cite{Artymowski:2017pua}.
We note that the value $0.234$ that normalizes Equation (\ref{Hbbn}) comes from the CMB data \cite{Planck:2018vyg}.
$\Delta N_{\nu,\text{eff}}=3.28-3.046$ is the difference between the cosmologically measured value and the SM prediction for the effective number of neutrinos after  $e^+ e^{-}$ annihilation  \cite{Cyburt:2004yc, Iocco:2008va}.
Apart from BBN, CMB anisotropy measurements also constrain the effective number of neutrinos
\cite{Lesgourgues:2013sjj}. Increasing the radiation density (or $\Delta N_{\nu,\text{eff}}$) at the time of recombination increases the angular scale of the photon diffusion length on the CMB sky and this reduces power in the damping tail of the temperature spectrum. Therefore, at the time of recombination it has to be $\rho_\text{GW}/\rho_\gamma \leq (7/8)\Delta N_{\nu,\text{eff}}$.
 
We now turn to gravitational radiation  that gives the stringent constraints to our scenario.

\section{Gravitational Waves and BBN/CMB Constraints}\label{S5}

There are two sources of primordial GWs that we are going to discuss. First, the quantum tensor perturbations generated during inflation, called inflationary or primary GWs \cite{Starobinsky:1979ty}. Second, the classical GWs generated by density perturbation, called induced or secondary GWs \cite{Matarrese:1992rp, Matarrese:1997ay} and for a recent review see \cite{Domenech:2021ztg}. There are also the Hawking radiated gravitons \cite{Anantua:2008am, Dong:2015yjs, Arbey:2021ysg,  Domenech:2021wkk}, as well as other aspects (see, e.g., \cite{Sandick:2021gew}),  whose implications are going to be examined in a separate work. 

\subsection{Reheating after a Kination Era}

Kination era follows inflation in our scenario. 
If $\beta<\beta_\text{thesh}$, Equation (\ref{thresh}), kination era ends by the Hawking radiation that dominates the energy budget of the universe.

\subsubsection{BBN/CMB Constraints on GWs from Inflation and a Kination Domination Phase}
A stringent constraint comes from the gravitational wave energy density, which is enhanced  in the GHz region during the kination regime \cite{Giovannini:1999bh, Riazuelo:2000fc, Yahiro:2001uh, Boyle:2007zx}.  
 The spectrum of GWs is described in terms of the fraction of their energy density per logarithmic wavenumber   interval $\rho_\text{crit}^{-1} d\rho_\text{GW}(\eta, k)/d\ln k$.
 In terms of the tensor power spectrum it is written as,
\begin{equation} \label{OmegaIGWc}
    \Omega_{\textrm{GW}}(\eta,k)\equiv{1\over24}\ \left({k\over\mathcal{H}(\eta)}\right)^2\ \overline{\mathcal{P}_{h}(\eta,k)},
\end{equation}
where ${\cal H}\equiv a^{-1}da/d\eta=2\eta^{-1}/(1+3w)$
is the conformal Hubble parameter and the overline denotes oscillation average. 
Assuming that the modes with wavenumber $k$  enter the horizon during radiation domination at the temperature $T_k$, the present-day value is
\begin{equation} \label{GWt0}
 \Omega_{\textrm{GW}}(\eta_0,k)\approx \Omega_\text{rad} (\eta_0)  {g_*(T_k) \over g_*(T_0)} \left({ g_{s}(T_0)\over g_{s}(T_k)}\right)^{4/3}   \Omega_{\text{GW}}(\eta_c,k),
\end{equation}
where $\eta_c$ denotes the conformal time that the generation of the primordial tensor modes has been completed.
We take the present day  thermal and entropy degrees of freedom to be $g_*(T_0)=3.36$ and $g_s(T_0)=3.93$, respectively, and equal at the high temperature $T_k$. The present value of the radiation energy density parameter is $h^2\Omega_\text{rad} (t_0)\approx 4.18 \times 10^{-5}$.
Detailed dependence on the degrees can be found in  \cite{Saikawa:2018rcs}.
Primordial tensors generated by quantum fluctuations during inflation are
\begin{equation} \label{Pt}
{\cal P}_h\equiv \frac{k^3}{2\pi^2} \sum_\lambda |h_{\textbf{k},\lambda}|^2=\frac{8}{M^2_\text{Pl}}\left(\frac{H_\text{inf}}{2\pi}\right)^2 \left(\frac{k}{aH}\right)^{n_t},
\end{equation}
where $H_\text{inf}$ is the Hubble parameter during inflation and $n_t$ is the tensor spectral tilt. On the other hand, the power spectrum of curvature
 perturbations is,
 \begin{equation}
 {\cal P}_{\cal R}
 \equiv \frac{k^3}{2\pi^2}  |{\cal R}_{\textbf{k}}|^2=\frac{1}{2\epsilon M^2_\text{Pl}}\left(\frac{H_\text{inf}}{2\pi}\right)^2 \left(\frac{k}{aH}\right)^{n_s-1}\,.
\end{equation}

Assuming that a scale $k$ enters the horizon in the early universe either during the early kination phase or during radiation domination that follows, and utilizing the ratio $ {\cal P}_h/ {\cal P}_{\cal R}=16\epsilon \equiv r$, the 
 energy density parameter of the inflationary GWs in the scales of interest is
\begin{align} \label{GWblue}
 \Omega_{\text{GW,inf}}(t_0,k)\approx 3\times 10^{-11}\alpha_k\, r_*\,\Omega_\text{rad}(t_0)  \left({g_*(T_0) \over 3.93} \right)^{4/3} \left(\frac{g_*(T_\text{rh})}{106.75}\right)^{-1/3}  \times 
\left\{ 
 \begin{array}{ll}
      k/k_\text{rh}\,, &  k > k_\text{rh}  \\
      1\,, &   k < k_\text{rh} \\
\end{array} 
\right. 
\end{align}
where we introduced the parameter $\alpha_k\equiv H^2_k/H^2_*$ to include the decrease of the Hubble parameter during inflation. In the above equation we took into account that gravitational radiation increases like $a^2$ during a kination phase and that the scale factor scans scales as $a \propto k^{-1/2}$ during that epoch.
 Following the discussion of the previous section, the energy density of GWs  that propagate inside the horizon at the time of big bang nucleosynthesis act as an additional radiation component increasing the background expansion rate, $3 H^2 M^2_\text{Pl}=\rho_\text{rad}+\rho_\text{GW}$.
 GWs produced prior to BBN
 do not alter BBN predictions  and do not change the position and amplitude of the CMB acoustic peaks 
only if 
$\rho_\text{GW}/\rho_\gamma \leq (7/8)\Delta N_{\nu,\text{eff}}$ at BBN and recombination eras, respectively
 \cite{Maggiore:1999vm, Caprini:2018mtu}.
This is equivalent 
to a bound on the integrated energy density, 
\begin{align} \label{GwBBN}
h^2 \int^{k_\text{end}}_{k_\text{m}} \Omega_\text{GW}(k,t_0)d \ln k & \leq 5.6 \times 10^{-6} \Delta N_{\nu,\text{eff}}   
\approx 1.3 \times 10^{-6}  \left( \frac{\Delta N_{\nu,\text{eff}}}{0.234} \right) \,.
\end{align} 

For the lower limit of the integral the scale $k_\text{m}=k_\text{CMB}\sim 10^{-2}$ Mpc$^{-1}$ is chosen,
that extends the integral to wider range of frequencies, compared to $k_\text{m}=k_\text{BBN}$,  giving a stringent constraint.
  An inflationary era is a quasi-de Sitter phase and it is roughly $H_\text{inf}\sim $ constant, slightly decreasing towards the end of inflation. This means that the tensor spectral tilt in Equation (\ref{Pt}) is $n_t\sim 0$. We can thus assume a roughly constant value $\Omega_\text{GW}(k, \eta_0)\approx\overline{\Omega}_\text{GW}\approx 8\times 10^{-16}\, r_*$, in the period between the moment $k_\text{CMB}^{-1}$  exits the horizon and the end of inflation.
In terms of inflationary quantities it is
 $k_\text{end}=k_* e^{N_*}(H_\text{end}/H_*)$. 
 $k_\text{end}$  can also be written in terms of reheating temperature and for a general equation of state according to the formula,
\begin{align} \label{kMgen}
k (M_\text{hor},  T_\text{rh}, &  w_\text{})  \simeq 2 \times 10^{17} \text{Mpc}^{-1} \left( \frac{T_\text{rh}}{10^{10}\text{GeV}} \right)^{\frac{1-3w}{3(1+w)}} 
\left(\frac{M_\text{hor}}{10^{12} \text{g}} \right)^{-\frac{3w+1}{3(1+w)}} \left(\frac{g_*}{106.75}\right)^{\frac{-2w}{6(1+w)}}.
\end{align}

For a kination domination regime that transits into radiation domination it is\linebreak $ k_\text{end}=k( M_\text{end}, T_\text{rh}, 1)$.
The fractional energy density  of GWs that impacts BBN is
\begin{align}
h^2 \int^{k_\text{end}}_{k_\text{CMB}} \Omega_\text{GW}(k,t_0)d \ln k &\approx
8\times 10^{-16}\, r_*\,
\left[ \left(\frac{k_\text{end}}{k_\text{rh}}\right) +\ln\left(\frac{k_\text{rh}}{k_\text{CMB}}\right) 
\right] \\
& \approx 8\times 10^{-16}\, r_*\, 
\left[ 5\times 10^{12}
\left(\frac{T_\text{rh}}{10^6 \,\text{GeV}} \right)^{-4/3}  
 \left(\frac{\alpha_\text{end} \,r_*}{10^{-2}} \right)^{1/3}
  \left(\frac{g_*}{106.75}\right)^{-1/6} \right. \nonumber  \\
 &\quad\quad\quad\quad\quad\quad\quad\quad\quad\quad\quad \left. + 35\ln\left( \frac{T_\text{rh}}{10^6 \text{GeV}}\right) \right]  \nonumber
\end{align}
where $\alpha_\text{end}\equiv H^2_\text{end}/H^2_* <1$
and we assumed a kination postinflationary phase until $t_\text{rh}$.
The CMB bound  (\ref{GwBBN}) implies that the reheating temperature for the kination domination models has to be larger than,
\begin{align} \label{TGW}
T_\text{rh} \gtrsim 10^{7}\, \text{GeV}\left( \frac{H_\text{end}}{H_*}\right)^{1/2}
\left(\frac{r_*}{10^{-2}} \right)
 \left( \frac{g_*}{106.75}\right)^{-1/8} 
 \left( \frac{\Delta N_{\nu,\text{eff}}}{0.234}\right)^{-3/4}\,.
\end{align}

  The reheating temperature for a kination phase  terminated by PBH evaporation   is given by  (\ref{Tkination}) and  a lower bound on the formation rate is obtained,
\begin{align} \label{betaBBN}
\beta \gtrsim  10^{-15}  \gamma^{-2} \left(\frac{r_*}{10^{-2}} \right)^{4/3} \left( \frac{\Delta N_{\nu,\text{eff}}}{0.234}\right)^{-1}\,, 
\end{align}
having  assumed that roughly  $H_\text{end}\sim 0.1 H_*$. Smaller values for $\beta$ mean smaller amounts of Hawking radiation and a delayed reheating 
of the universe that experiences an extended kination domination phase.   
Comparing with $\beta_\text{thresh}$ (\ref{thresh}) we see that it is $\beta<\beta_\text{thresh}$ only for small PBH masses or for small $r_*$ values.
 The bound (\ref{betaBBN}) is relaxed if  we add an initial amount of radiation in the postinflationary universe, due to, e.g., gravitational reheating, see Appendix \ref{A1}.

The aferomentioned results are modified if  we depart from the simplistic approximation of $w=1$. In that case  an effective average equation of state during the entire reheating period should be considered,
\begin{equation}
\bar{w}_\text{rh}=\frac{1}{N_\text{rh}}\int^{N_\text{rh}} w_\text{rh}(N)dN=\frac{1}{\ln\left(\frac{a_\text{rh}}{a_\text{end}}\right)} \int^{a_\text{rh}}_{a_\text{end}} w_\text{rh}(a)\frac{da}{a}\,.
\end{equation}

\subsubsection{BBN/CMB Constraints on Induced GWs from Kination Era}

The induced GWs are sourced gravitational waves. 
The leading source is the scalar perturbations that evolve according to the equation,  
\begin{align}
\Phi''(x)+\frac{6(1+w)}{1+3w}\frac{1}{x}\Phi'(x)+w\Phi(x)=0\,.
\end{align}
Here we have implicitly assumed an negligible anisotropic stress and introduced the variable  $x\equiv k\eta$.
$\Phi$ is the scalar transfer function defined $\Phi_{\textbf{k}}(\eta)\equiv \Phi(x)\ \phi_{\textbf{k}}$ where $\phi_{\textbf{k}}$ is the Fourier mode of the primordial scalar perturbations and $\Phi_{\textbf{k}}$ of the  Bardeen potential.
For the kination domination  scenario, $w=1$, it is $\Phi(x)=\frac{2}{x}J_1(x)$, and if we consider a  transition  into RD  the scalar transfer function is given by \cite{Dalianis:2020cla},
\begin{equation}
\label{pot}
    \Phi(x)=\begin{cases}
    \displaystyle
    {2\over x}\ J_{1}(x), & x<x_\text{rh}\\
    \displaystyle
    {3\over x^2}\left[ C_1 \left({\sin(x/\sqrt{3})\over x/\sqrt{3}} - \cos(x/\sqrt{3})\right) + C_2 \left({\cos(x/\sqrt{3})\over x/\sqrt{3}} + \sin(x/\sqrt{3})\right) \right], & x\ge x_\text{rh}
    \end{cases}
\end{equation}
 where $x_\text{rh}$ corresponds to the 
moment of reheating.
 The coefficients  $C_1$ and $C_2$ are determined by the continuity of the potential and its derivative at the point of the transition, after expressing $J_{3/2}$ and $Y_{3/2}$ in terms of spherical Bessel functions.
$\Phi(x)$ remains constant as long as $x\ll 1$. 
According to  Equation~(\ref{pot})   the gravitational potential starts decaying at the horizon crossing roughly as $x^{-3/2}$ experiencing maximal pressure during  KD. After the transition to the RD phase the decay is even faster, $x^{-2}$.
At some time  $t_\text{c}$, that might be during the kination or radiation era, the source $\Phi(x)$ has decayed and the production of IGWs~ceases. 

The power spectrum of the induced gravitational waves is expressed as a double integral of  the curvature power spectrum and a tensor transfer function,
\begin{equation}
\label{tensorPSD}
    \overline{\mathcal{P}_{h}(\eta,k)}=\int_{0}^{\infty} dv\int_{|1-v|}^{1+v} du\ \mathcal{T}(u,v,\eta,k)\ \mathcal{P}_{\mathcal{R}}(u
k)\ \mathcal{P}_{\mathcal{R}}(v k).
\end{equation} 
where the over-line denotes the oscillation average. 
The tensor transfer function is given by
\begin{align} \label{kernel}
    \mathcal{T}(u,v,\eta,k)=4\ \left(4 v^2-(1+v^2-u^2)^2\over4 u v\right)^2
    \left({3+3w\over5+3w}\right)^2\ \overline{I^2(u,v,\eta,k)},
\end{align}
where   $ \overline{I(u,v,\eta,k)}$
is  the oscillation average of a kernel function
 composed of the Green's function  $kG_{k}(x,y)={\pi\over2} \sqrt{x\ y}\cdot \left[Y_{\nu}(x)\ J_{\nu}(y)-Y_{\nu}(y)\ J_{\nu}(x)\right]$, with $\nu\equiv{3(1-w)\over2(1+3 w)}$. 
For kination domination it is $\nu=0$ and the Bessel functions $J_0$ and $J_1$ do not have a closed form representation and  can only be written as infinite series. Therefore the transfer function $\mathcal{T}$ in kination domination, contrary to radiation domination,
cannot be written in closed analytical expressions and the computation has to be conducted numerically.

For the scenario of the kination domination that transits into radiation  and for  a monochromatic power spectrum of scalar perturbations modeled by a delta-distribution peaked at $k_\text{p}$,  $\delta(k)=A_0 \delta(\ln(k/k_\text{p}))$, the  energy density parameter  at the moment $t_\text{c}$ is~\cite{Domenech:2019quo, Dalianis:2020cla}
\begin{align} \label{IGWkin}
    \Omega_{\textrm{IGW}}^{\textrm{KIN}\rightarrow \textrm{RD}}(\eta_\text{c},k) &={1\over6}A_{0}^2\ \left({k_\text{p}\over\mathcal{H}(\eta_\text{c})}\right)^2\ \left[1-\left(k\over2 k_\text{p}\right)^2\right]^2
    \overline{I_{\textrm{KIN}\rightarrow\textrm{RD}}^2\left({k_\text{p}\over k},{k_\text{p}\over k},\eta_\text{c},k\right)}\ \Theta\left(1-{k\over2 k_\text{p}}\right),
\end{align}
where the unit step function $\Theta$ is for the conservation of momentum cutting off  tensor modes with $k>2k_\text{p}$. For $\eta_c \ll \eta_\text{rh}$ no sharp peak in the IGWs appears even for a monochromatic ${\cal P_R}(k)$.  
After the moment $t_c$  the GWs propagate freely. The present-day  energy density parameter of the induced GWs is given by Equation (\ref{GWt0}).
 The energy density of induced GWs for a broad spectrum of curvature perturbations can be found from the expression (\ref{IGWkin}) by the correspondence 
$A_0 \longleftrightarrow A_{\cal R}\equiv {A_0}/(\epsilon \sqrt{\pi})$ 
 where $\epsilon$ is the width of a  Gaussian distribution\cite{Dalianis:2020cla}.

Similarly here, during kination domination the growth of induced GWs relative to the background is proportional to $a^2$, given by the expression (\ref{GWblue}).  In terms of conformal time it is
\begin{align}
\Omega_{\textrm{IGW}}^{\textrm{(KIN)}}\sim (k_\text{p}\eta)^2\overline{I^2_{\textrm{KIN}}}
\propto\eta\propto a^2\,, \quad\text{for}\,\,\, \eta<\eta_\text{rh}
\end{align}

This growth stops after the transition to the radiation domination phase.
If it is  $\eta_\text{entry}\ll \eta_\text{rh}$ we can neglect the effect from the RD era on the shaping of induced GWs, because by that time the gravitational potential sourcing the tensor modes is negligibly small.
In the case of induced GWs, it is much different from the GWs produced by the de-Sitter stage of inflation. The spectrum of induced GWs has a prominent peak at wavenumbers  relevant to PBH production. If there was not this amplification of the scalar power spectrum, the induced GWs would be irrelevant for our discussion.   Notwithstanding,   the existence of an early kination phase implies that the energy density of the produced induced GWs, associated with PBH production, becomes enhanced and might backreact on the geometry. 

The induced GWs do not spoil BBN/CMB if they satisfy the bound (\ref{GwBBN}). 
In particular, the  increase of the expansion rate does not alter the position and amplitudes of the acoustic  peaks of CMB if
\begin{align} \label{I1BBN}
 h^2 \int^{k_\text{end}}_{k_\text{rh}} 
\frac{k}{k_\text{rh}}\Omega_\text{IGW}(k, \eta_0)d \ln k + 
h^2 \int^{k_\text{rh}}_{k_\text{CMB}} 
\Omega_\text{IGW}(k, \eta_0)d \ln k
\leq 1.3 \times 10^{-6}  \left( \frac{\Delta N_{\nu,\text{eff}}}{0.234} \right) \,.
\end{align} 

A numerical integration of a spectrum of induced GWs with peak at $k\sim 10^{22}$ Mpc$^{-1}$ and reheating temperature $T_\text{rh}\sim 10^9$ GeV associated with PBHs with mass $M_\text{PBH}\sim 10^5$~g, generated by a sharp peak at the scalar spectrum,  violates the above bound five orders of magnitude.
In order to perform the integration and demonstrate the conflict of induced GWs with BBN predictions, we will assume a rough, top-hat approximation about the peak  for the spectrum of induced GWs. 
Let us assume that the energy of GWs is stored mainly in a narrow wave band $(k_1, k_2)$ with central wavenumber $k_\text{p}$ that enters the horizon at $\eta_\text{entry}$. 
The  top-hat approximation for the GW spectrum with amplitude $\Omega_\text{IGW}=A_\text{IGW}$ in the interval $(k_1, k_2)$ around $k_p$ gives,
\begin{align} \label{IGWkinBBN}
\frac{k_\text{p}}{k_\text{rh}} 
\,A_\text{IGW} \,
\ln\left(\frac{k_2}{k_1}\right)
\lesssim 0.1 \left( \frac{\Delta N_{\nu,\text{eff}}}{0.234} \right) \,.
\end{align} 

$A_\text{IGW}$ is related to the scalar power spectrum amplitude and we can consider that $\Omega_\text{GW}(k_\text{p})\sim 10^{-2}A_{\cal R}^2$, justified by our analytic and  numerical study, where $A_{\cal R}\equiv {\cal P_R}(k_\text{p})$. Assuming a  width $k_2/k_1\sim 10$,
we can express the above bound in terms of the reheating temperature,
\begin{align} \label{TIGW}
   T_\text{rh}\gtrsim  10^{10}\,{\text{GeV}} \,
   \left(\frac{A_{\cal R}}{10^{-2}}\right)^{3/2}\left(\frac{M_\text{PBH}/\gamma}{10^{5}\text{g}} \right)^{-1/2} 
\left(\frac{g_*}{106.75}\right)^{-1/8}    
   \left( \frac{\Delta N_{\nu,\text{eff}}}{0.234} \right)^{-3/4}
\end{align}
where we used that  $k_\text{rh}\approx 2\times 10^7 (T_\text{rh}/\text{GeV}) \, \text{Mpc}^{-1}$ and Equation (\ref{kMgen}) to express the wavenumbers.
This bound is much stringent than the bound coming from the inflationary GWs, Equation (\ref{TGW}).  
For reheating following a kination domination phase 
$\beta$ has to be larger~than   
\begin{align}
  \beta\gtrsim  10^{-11}\, \gamma^{-2}
   \left(\frac{A_{\cal R}}{10^{-2}}\right)^{2}\left(\frac{M_\text{PBH}/\gamma}{10^{5}\text{g}} \right)^{-2/3} 
\left(\frac{g_*}{106.75}\right)^{-1/6}    
   \left( \frac{\Delta N_{\nu,\text{eff}}}{0.234} \right)
   \end{align}
and at the same time less than $\beta_\text{thres}$. 
This happens only for too small $M_\text{PBH}$ masses, less than 0.1 grams, which contradicts the bound (\ref{minMpbh}). In other words, $\beta$ has to be large enough so that it is always $\beta>\beta_\text{thesh}$ and a PBH domination phase is unavoidably realized.

\subsection{Reheating after a PBH Domination Era}\label{S5.2}

For a PBH domination phase the above bounds alter significantly.
  PBH domination is realized for $\beta>\beta_\text{thresh}$ 
and such values are obtained for a slight increase of the curvature power spectrum, see Equation (\ref{brad}). 
A PBH domination phase has been discussed in different contexts and has profound implications on GW signals, see, e.g., \cite{Anantua:2008am, Zagorac:2019ekv, Inomata:2020lmk, Papanikolaou:2020qtd, Domenech:2020ssp}. In the following we will mainly discuss the evident impact of a matter domination phase on the GW energy density parameter, which is a $a^{-1}$ decrease,  considering inflationary and induced GWs.

\subsubsection{BBN/CMB Constraints on GWs from Inflation and a PBH Domination Phase}

A PBH domination era suppresses the energy density of  GWs with respect to the background.
The reheating temperature of a PBH domination phase is larger than that of a kination phase for the same PBH mass. Moreover, after inflation,  the pre-radiation  phase is partitioned between kination and PBH domination, see Figure \ref{Fig02bounds}.

 The ratio of the scale factor at  BBN and at the end of inflation is analyzed as
\begin{align}
\frac{a_\text{end}}{a_\text{eMD}}\frac{a_\text{eMD}}{a_\text{rh}}\frac{a_\text{rh}}{a_\text{BBN}}=e^{\left(N_\text{KIN}+N_\text{eMD} +N_\text{eRD} \right)}\,,
\end{align}
where kination domination lasts $N_\text{KIN}\equiv \ln(a_\text{end}/a_\text{eMD})$ efolds, PBH domination (early matter domination) lasts
$N_\text{eMD}\equiv \ln(a_\text{eMD}/a_\text{rh})$
efolds, and a pre-BBN radiation domination $N_\text{eRD}\equiv \ln(a_\text{rh}/a_\text{BBN})$ efolds.
Accordingly, the GW energy density during  BBN  is decomposed as
\begin{align} \label{integr}
\left(\frac{k_\text{rh}}{k_\text{eMD}} \right)^2\int^{k_\text{end}}_{k_\text{eMD}} 
\frac{k}{k_\text{eMD}}\Omega_\text{GW}(k, \eta_0)d \ln k + 
\int^{k_\text{eMD}}_{k_\text{rh}}\left( \frac{k_\text{rh}}{k}\right)^2 
\Omega_\text{GW}(k, \eta_0)d \ln k+
 \int^{k_\text{rh}}_{k_\text{BBN}} 
\Omega_\text{GW}(k, \eta_0)d \ln k
\end{align}

For inflationary GWs, that are roughly scale invariant, $\Omega_\text{GW}(k, \eta_0)\approx\overline{\Omega}_\text{GW}$,
the above expression is written in terms of efolds, 
\begin{align}
\overline{\Omega}_\text{GW} h^{2} & \left(\frac{e^{2N_\text{KIN}}-1}{e^{N_\text{eMD}}} +\frac12\left(1-e^{-N_\text{eMD}} \right) +
 N_\text{eRD}\right)  \approx \\
& \approx 8\times 10^{-16}\, r_*\left(\frac{e^{2N_\text{KIN}}}{e^{N_\text{eMD}}} + \ln\left(10^4\frac{T_\text{rh}}{\text{GeV}}\right)\right)\lesssim 1.3 \times 10^{-6}  \left( \frac{\Delta N_{\nu,\text{eff}}}{0.234} \right)\,, \nonumber
\end{align}
and the BBN bound is satisfied for 
\begin{align} \label{NkinB}
N_\text{KIN} \lesssim 12+\frac{N_\text{eMD}}{2}+\frac{1}{2}\ln\left(\frac{10^{-2}}{r_*}\right) +\frac12\ln\left( \frac{\Delta N_{\nu,\text{eff}}}{0.234} \right)\,.
\end{align}

PBHs form during kination at  $t_\text{form}$ about $N_\text{form}$ efolds after inflation, 
\begin{align} 
N_\text{form}=\frac{1}{2} \ln\left(\frac{k_\text{end}}{k_\text{form}} \right)=\frac{1}{3} \ln\left(\frac{M_\text{PBH}/\gamma}{M_\text{end}} \right)\,.
\end{align}

For $M_\text{PBH}\sim 10^5$ g it is $N_\text{form}\sim  4$ and for $M_\text{PBH}\sim 10^9$ g it is $N_\text{form}\sim 7$.
PBHs dominate the energy budget of the early universe after $\Delta N_\text{dom}$ efolds, 
\begin{align}
\Delta N_\text{dom}\approx-\frac{1}{3}\ln(\gamma \beta),
\end{align}
and it is $N_\text{KIN}=N_\text{form}+\Delta N_\text{dom}$. PBH domination lasts $N_\text{eMD}$ folds until reheating that coincides with PBH evaporation. It is $a_\text{eMD}/a_\text{form}\sim (t_\text{eMD}/t_\text{form})^{1/3}=(\gamma \beta)^{-1/3}$, $a_\text{rh}/a_\text{eMD}\sim (t_\text{evap}/t_\text{eMD})^{2/3}$ and 
\begin{align}
N_\text{eMD}\approx \frac{2}{3}\ln\left(\frac{M^3_\text{PBH} \gamma\beta}{e^{3N_\text{form}}H_\text{end}^{-1}m^4_\text{Pl}} \right)
=\frac{2}{3}\ln\left(\frac{M^2_\text{PBH} \gamma^2\beta}{2m^2_\text{Pl}} \right)
\end{align}

Summing up the expressions for the amount of efolds, the bound on kination era (\ref{NkinB}) is recast into a bound on $\beta$ value,
\begin{align} \label{gwInfpbh}
 \beta \gtrsim 10^{-16} \,\gamma^{-2} \left(\frac{H_\text{end}}{H_*}\right)^2 \left( \frac{M_\text{PBH}}{10^{5}\text{g}}\right)^{-1/2} \left( \frac{r_*}{10^{-2}}\right)\left( \frac{\Delta N_{\nu,\text{eff}}}{0.234} \right)^{-1/2}
\end{align}

This bound selects a wide part of the PBH domination $(M_\text{PBH}, \beta)$ parameter space where the tilted GWs from inflation do not change BBN predictions.
Note that the lower limit of integration interval (\ref{integr}) $[k_\text{BBN}, k_\text{rh}]$ slightly modifies the result, thus the above bound applies also if it is extended to the CMB scale, $[k_\text{CMB}, k_\text{rh}]$.

\subsubsection{BBN/CMB Constraints on Induced GWs from Kination Era  and  a PBH Domination~Phase}
 
\begin{figure}[!htbp]
  \includegraphics[width=0.49\linewidth]{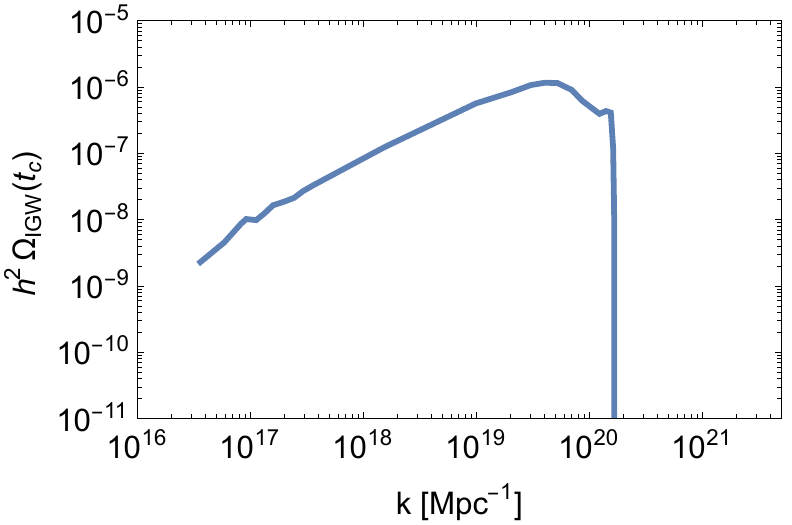}
  \includegraphics[width=0.49\linewidth]{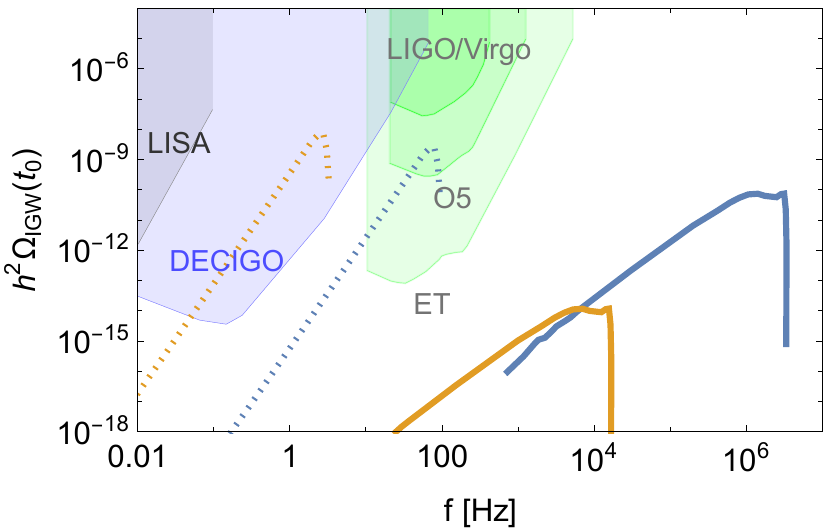}
\caption{ In the \textbf{left panel} the early spectrum of induced GWs  is depicted, produced during a kination domination phase by a monochromatic ${\cal P_R}(k)$ and associated with PBHs with mass  $M_\text{PBH}=10^5$ g. In the \textbf{right panel} the present day frequency spectrum is depicted for $M_\text{PBH}=10^5$ g (blue curve) and  $M_\text{PBH}=10^8$ g (orange curve) and for $\beta\sim 10^{-10}$.  Note the change of the shape of induced GW spectrum due to kination and early matter domination phase. The dotted curves depict the corresponding  isocurvature-induced GWs, following Ref.  \cite{Domenech:2020ssp}. 
The colored areas are enclosed by the sensitivity curves of LIGO/Virgo, Einstein Telescope, LISA and DECIGO.
} \label{FigIGW}
\end{figure}

The early matter domination era realized by PBHs suppresses the relative enhancement of induced GWs relative to the background energy density, see Figure \ref{FigIGW}. 
For the simplistic top-hat approximation for the energy density spectrum of induced GWs the suppression is given by the exponential in front of the brackets,
\begin{align} \label{IGWdom}
e^{{-N_\text{eMD}}+2\Delta N_\text{dom}}\left[\frac{k_\text{p}}{k_\text{rh}} 
\,A_\text{IGW} \,
\ln\left(\frac{k_2}{k_1}\right)\right]
\lesssim 0.1 \left( \frac{\Delta N_{\nu,\text{eff}}}{0.234} \right) \,.
\end{align} 
where we considered modes that propagate inside the horizon at the time of BBN/CMB.
 Induced GWs, associated with the PBH formation, do not change BBN predictions for 
 \begin{align} \label{masterbound}
 \beta \gtrsim 10^{-13} \,\gamma^{-3/2} \left(\frac{A_{\cal R}}{10^{-2}}\right)^{3/2} 
 \left( \frac{M_\text{PBH}}{10^{5}\text{g}}\right)^{-1} \left( \frac{\Delta N_{\nu,\text{eff}}}{0.234} \right)^{-3/4}\,.
 \end{align}
 
\begin{figure} [!htbp]
  \includegraphics[width=.49 \linewidth]{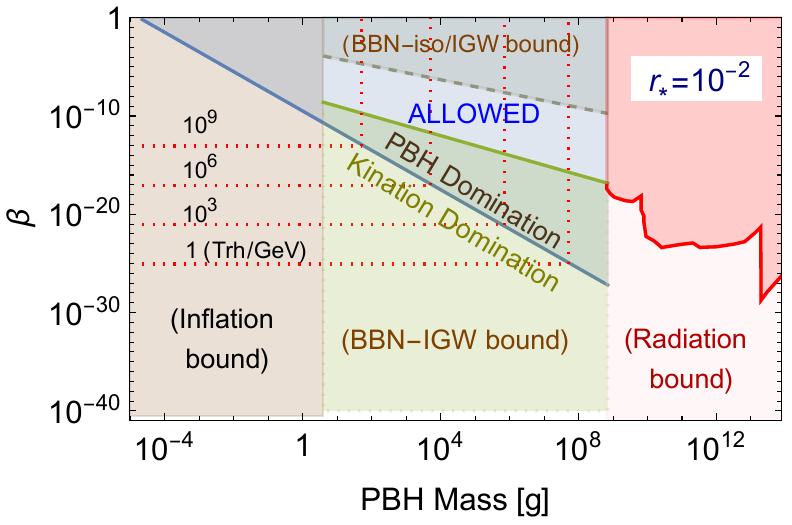}
  \includegraphics[width=.49 \linewidth]{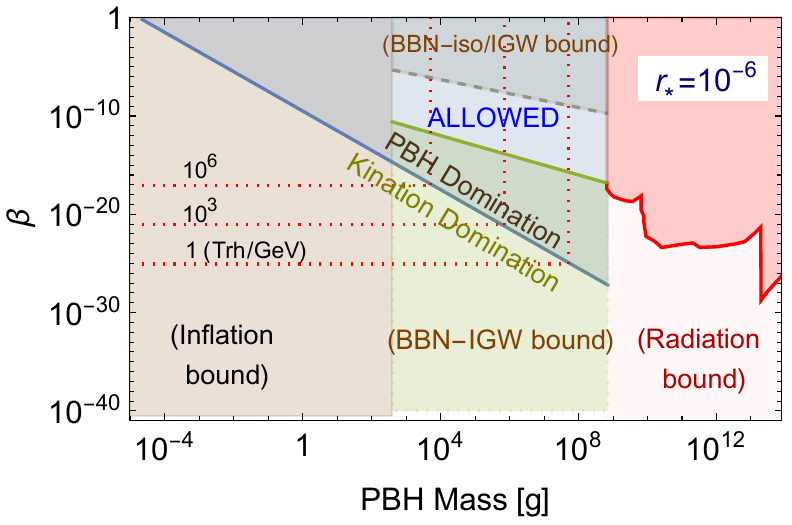}
\caption{In this figure we depict  the combined constraints for the scenario of an early kination domination phase reheated  via PBH evaporation.
 The allowed 
 parameter space ($M_\text{PBH}, \beta$), that  gives viable cosmology, lies in the PBH domination region.
The \textbf{left} part of the parameter space (gray area) is excluded 
by the energy scale of inflation that determines the minimum PBH mass, shown  for two different  $r_*$ values. The \textbf{right} part of the parameter space is excluded by the requirement of radiation domination during BBN (red area).  
 The induced GW constraints due to BBN/CMB
 exclude a great part (green area) of the parameter space as described in the text.  
 The red dotted lines indicate the reheating temperature.
} \label{FigAll}
\end{figure}

This is the most stringent constraint posed on the scenario of runaway inflation  reheated via PBH evaporation.
We illustrate this constraint in 
 \ref{FigAll}.
Note that the constraint (\ref{masterbound}) encompasses the 
constraint (\ref{gwInfpbh}) from inflationary GWs.

\subsubsection*{Induced GWs and Detection by LIGO/Virgo, Einstein Telescope}

The induced GWs associated with the production of mini PBHs have large frequencies that can be probed only from LIGO/Virgo experiment and the designed  Einstein Telescope.   Focusing on the frequency band around  $100$ Hz, which corresponds to wavenumbers about $k_\text{EX}\sim 10^{16}$ Mpc$^{-1}$, the induced GWs must have an amplitude $\Omega(t_0, k_\text{EX})$ below the sensitivity curve of the operating experiments. It is exciting that current or future GW experiments can rule out or support this cosmological scenario, probing part of its parameter space, see Figure~\ref{FigIGW}.

\subsubsection*{Isocurvature-Induced GWs}

Ref. \cite{Papanikolaou:2020qtd} noted that the isocurvature perturbation of PBHs (and of dark matter in general \cite{Domenech:2021and}) induces additional tensor modes.  As discussed in Section \ref{S2}, PBH formation is a rare event and takes place at those regions of space where the density perturbation is above the threshold $\delta_c$. Their space distribution is random and follows a Poisson type distribution~\cite{Meszaros:1975ef, Afshordi:2003zb, Papanikolaou:2020qtd}. 
For distances much larger than their mean separation, $\bar{r}=((3/4\pi) n^{-1}_\text{PBH})^{1/3}$, the PBH gas can be described as a continuous fluid that acts as an isocurvature perturbation $S=3(\zeta_\text{PBH}-\zeta_\varphi)$ where $\zeta=-\Phi+H\delta \rho/\dot{\rho}$ is the curvature perturbation on uniform-energy density hypersurfaces and $\Phi$ is the Bardeen potential \cite{Wands:2000dp}. For the kination and PBH fluid they, respectively,   $\zeta_\varphi=-\Phi+(1/6)\delta_\rho\varphi/\rho_\varphi$, $\zeta_\text{PBH}=-\Phi+(1/3)\delta \rho_\text{PBH}/\rho_\text{PBH}$, 
thus, at the formation time and with $\beta\ll 1$ it is
\begin{align}
S=\frac{\delta \rho_\text{PBH}}{\rho_\text{PBH}} -\frac12 \frac{\delta\rho_\varphi}{\rho_\varphi} \approx \frac{\delta \rho_\text{PBH}}{\rho_\text{PBH}}
\,,
\end{align}
where the second equality follows from the isocurvature property $\delta\rho_\varphi+\delta\rho_\text{PBH}=0$.
$S$ and separately $\zeta_\varphi$, $\zeta_\text{PBH}$ are conserved 
on superhorizon scales.
Initially the curvature perturbation receives its leading contribution from the scalar field $\zeta\approx \zeta_\varphi$, it evolves with time and in the PBH domination phase it is $\zeta=\zeta_\text{PBH} \approx \zeta_\varphi+S/3$.
The comoving curvature perturbation ${\cal R}$ is related to $\zeta$ by a gauge transformation
and coincide in the limit $k\rightarrow 0$. During PBH domination it is ${\cal R}=-\zeta \approx 5\Phi/3$.
The evolution of the Bardeen potential is given by the equation
\begin{align}
\Phi''+3{\cal H}(1+c_s^2)\Phi'+\left({\cal H}^2(1+3c^2_s) +2{\cal H}'\right) -c^2_s \nabla^2 \Phi=\frac{a^2}{2} \rho_\text{PBH} c^2_s S
\end{align}
where the sound speed is 
\begin{align}
c^2_s\equiv \frac{\rho_\text{rad}}{\rho_\text{rad} +\frac12\rho_\text{PBH}}
\end{align}

The early isocurvature sources tensor modes that have size given by the formula (\ref{tensorPSD}) after replacing ${\cal P_R}\rightarrow {\cal P}_S$, where \cite{Papanikolaou:2020qtd, Domenech:2020ssp}
\begin{align}
{\cal P}_S(k)=\frac{2}{3\pi}\left(\frac{k}{k_\text{UV}}\right)^3
\end{align}
is the initial isocurvature perturbation power spectrum. $k_\text{UV}=a/\bar{r}$ is an ultra-violet cutoff below which the grainy structure of the PBH fluid becomes important. 

According to \cite{Inomata:2019ivs, Inomata:2019zqy}, and applying the notation to our scenario that discusses a kination era, the Kernel $I(u,v,\eta, k)$ (\ref{kernel}) is split into three contributions, $I=I_\text{KIN}+I_\text{eMD}+I_\text{RD}$, from kination era, PBH domination and radiation era, respectively. 
$I_\text{RD}$ gives the dominant contribution to isocurvature induced GWs due to non-zero potential $\Phi$ field  at small subhorizon scales. We recall that
the perturbation $\Phi$ decays during kination domination but it remains constant during PBH domination and decays completely only after reheating of the universe. Ref. \cite{Domenech:2020ssp} analyzed a similar to our case problem with the difference that an early radiation era precedes PBH domination. They find a peak of the  isocurvature induced GWs around $k\sim k_\text{UV}$ with amplitude that depends on $\beta^{16/3}$ and $M_\text{PBH}^{34/9}$. Taking into account the BBN requirements $\beta$ value is restricted in the range \cite{Domenech:2020ssp},
\begin{align}
6 \times 10^{-9} \left(\frac{M_\text{PBH}}{10^5}\right)^{-1}\lesssim 
\beta \lesssim 2\times 10^{-6} \left( \frac{M_\text{PBH}}{10^5 \text{g}}\right)^{-17/24}
\end{align}

As a rule of thumb we can apply the above bounds to kination domination scenario, see Figure \ref{FigAll}, because isocurvature-induced GW receive their main contribution at the late stages of PBH domination and, particularly, at the beginning of radiation domination.    We remark that the sudden transition \cite{Inomata:2019ivs} from PBH domination to radiation, that gives the maximal GW and detectable signal, Figure \ref{FigIGW}, is realized only for a universal sudden evaporation that is possible only for a  monochromatic PBH mass spectrum. Moreover, as noted in \cite{Papanikolaou:2020qtd, Domenech:2020ssp}, the density contrast of PBH fluctuations exceeds unity at the time of evaporation and the linear analysis should break down at some point. A non-perturbative methodology
 has been developed in Ref. \cite{Dalianis:2020gup} where the deviations
 from ashpericity due to the absence o pressure has been explicitly considered. An overdensity of size $k^{-1}$ evolves non-spherically and eventually collapses at a pancake configuration at the time $t_\text{col}$.
  The spectral energy density parameter is given in terms of the quadrupole tensor $Q_{ij}$  
 and a probability density function ${\cal F_D}(\alpha, \beta, \gamma)$, that describes the size of asphericity of the~overdensity, 
 \begin{align} 
\Omega_\text{GW}(t_0,f_0) \propto  G
\int\int\int
\,d\alpha d\beta d\gamma\, |\tilde{Q}_{ij}(f)|^2 
{{\cal F}_\text{D} (\alpha, \beta,\gamma, \sigma)}\,,  \label{result_Om}
\end{align} 
with the precise expression found in Ref. \cite{Dalianis:2020gup}.
We leave a detailed analysis of this particular problem for a future work.

\section{PBHs from  Runaway Inflation Models }\label{S6}

The inflationary potential that we consider is a runaway one without a minimum.
Its precise form plays a critical  role for the determination of both  the inflationary  and postinflationary cosmic evolution.
The number of efolds $N_*$ has to be large, compatible with a kination domination phase that follows inflation.
Furthermore,  the position and characteristics of the feature  that amplifies the power spectrum specify the mass and the abundance of the PBHs.  
A runaway  potential is acceptable only if it can realize  a sufficient reheating of the universe and, here, this is realized by the production and evaporation of mini PBHs.

  The construction of runaway  inflation models that induce PBH production and act as quintessence models today is very challenging.
This unified description works only for particular choices of the parameters that affect horizontally  the observables in the cosmic time range from  $10^{-35}$ s to 13.8 billion years.
To be specific, we mention that the number of efolds $N_*$, related to the CMB scale, depends on 
the reheating temperature which is not a free parameter but depends on  $\beta$ and $M_\text{PBH}$. This is in sharp contrast with conventional  inflation models  where the inflaton field decays perturbatively about the vacuum.
A particular PBH mass $M_\text{PBH}$ is related to a specific $k$, where ${\cal P_R}(k)$ is amplified,  only after  $N_\text{KIN}$, $N_\text{eMD}$ are  known.
Hence, the characteristics of the  peak in the power spectrum determine
 the mass of the evaporating PBHs and the reheating temperature of the universe and the remnant abundance!
 Additionally, the tail of the potential might lead to the observed late time acceleration of the universe.

In single field models, the special feature that amplifies the power spectrum can be either an inflection point \cite{Garcia-Bellido:2017mdw, Motohashi:2017kbs, Ballesteros:2017fsr}  or a step-like transition in the inflationary plateau \cite{Kefala:2020xsx, Dalianis:2021iig}\footnote{Amplification of the power spectrum due to non-minimal coupling of the inflaton to gravity has been also proposed \cite{ Dalianis:2019vit, Fu:2019ttf}.}. 
The presence of step-like, or generally sharp, features in the inflation potential results in amplification and, additionally, in oscillatory patterns in  the curvature power spectrum \cite{Dalianis:2021iig, Fumagalli:2020nvq}. The potential might feature more than  one steps, so that the relevant enhancement in the power spectrum becomes strong enough. 

In the following subsections we will discuss in some detail the inflection point scenario.

\subsection{Building a ${\cal P_R}(k)$ Peak}

The PBH abundance is found  after  computing  the value of the comoving curvature perturbation ${\cal R}_k$. 
 In the comoving gauge we have  $\delta \varphi=0$ and $g_{ij}=a^2 \left[(1-2{\cal R}) \delta_{ij} +h_{ij}\right]$, Expanding the inflaton-gravity action
to second order in ${\cal R}$ one obtains
\begin{equation}
S_{(2)}= \frac12 \int {\rm d}^4x \sqrt{-g} a^3 \frac{\dot{\varphi}^2}{H^2} \left[\dot{{\cal R}}^2 -\frac{(\partial_i {\cal R})^2}{a^2}\right]\,.
\end{equation} 

We can write this action in terms of the variable  $v=z{\cal R}$, where $z^2=a^2\dot{\phi}^2/H^2=2a^2\epsilon_1$.  
The evolution of the Fourier modes $v_k$ is given by the Mukhanov--Sasaki equation
\begin{equation} \label{MS}
v''_k +\left(k^2-\frac{z''}{z}\right) v_k =0 ,
\end{equation}
where we switched to conformal time $\eta$ and $z''/z$ is  expressed in terms of the Hubble-flow functions  
\begin{equation} \label{eek}
\epsilon_1 \equiv -\frac{\dot{H}}{H^2}, \quad
 \epsilon_2 \equiv \frac{\dot{\epsilon}_1}{H\epsilon_1}, \quad 
 \epsilon_3 \equiv \frac{{\dot{\epsilon}}_2}{H\epsilon_2}, 
\end{equation}
as 
\begin{equation}
\frac{z''}{z} =(aH)^2 \left[2-\epsilon_1 +\frac32 \epsilon_2 - \frac12\epsilon_1 \epsilon_2 +\frac14 \epsilon^2_2+\frac{1}{2} \epsilon_2 \epsilon_3 \right]. 
\end{equation}

\begin{figure}[!htbp]
  \includegraphics[width=0.518\linewidth]{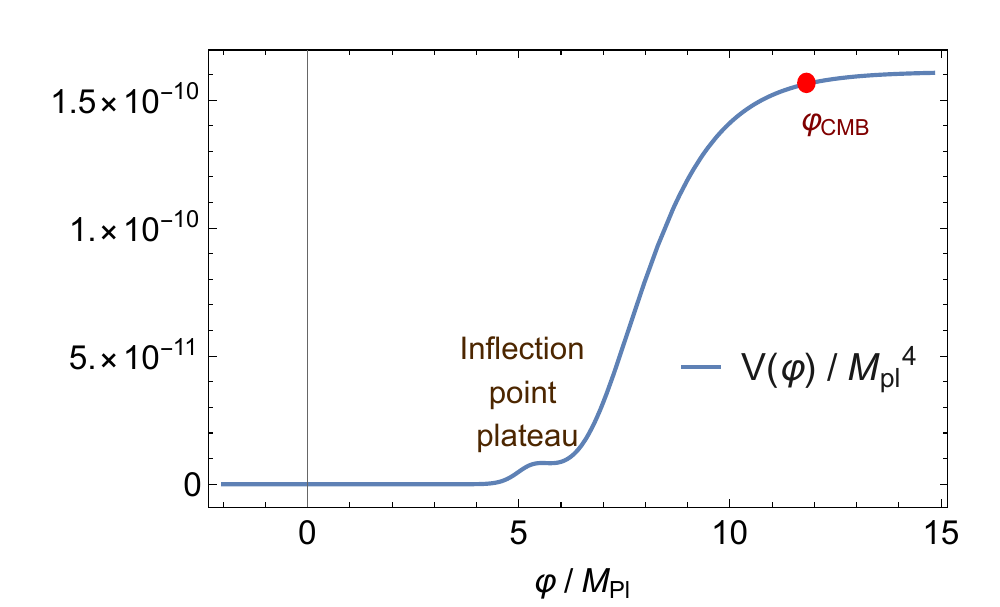}
  \includegraphics[width=0.49\linewidth]{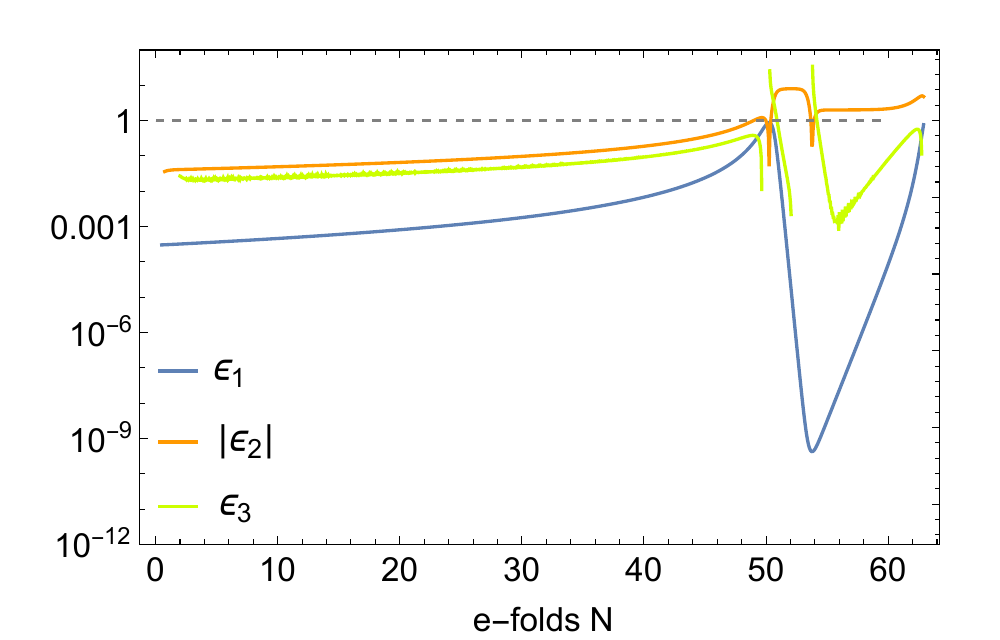}
\caption{In the \textbf{left panel} the potential of a  runaway inflaton field described by superconformal attractors is depicted.
The inflaton field value corresponding to the CMB scale $k=0.05$ Mpc$^{-1}$ is denoted with a red dot.
 In the \textbf{right panel} the Hubble flow parameters $\epsilon_1$, $\epsilon_2$,  $\epsilon_3$ with respect to the e-folds number   
  are depicted  demonstrating the slow-roll violation that enhances the power spectrum at  the end of the inflationary stage.
} \label{FigVSR}
\end{figure}

The evolution of the Hubble-flow functions  for a runaway potential with an inflection point is depicted in Figure \ref{FigVSR}.
The power of ${\cal R}_k$ 
 estimated at a time well after the mode exits the horizon and its value freezes out is
\begin{equation}
 \left. {\cal P_ R} =\frac{k^3}{2\pi^2}\frac{|v_k|^2}{z^2} \right|_{k\ll aH } \,.  \label{ppp}
\end{equation}

\begin{figure}[!htbp]

\includegraphics[width = 0.6\textwidth]{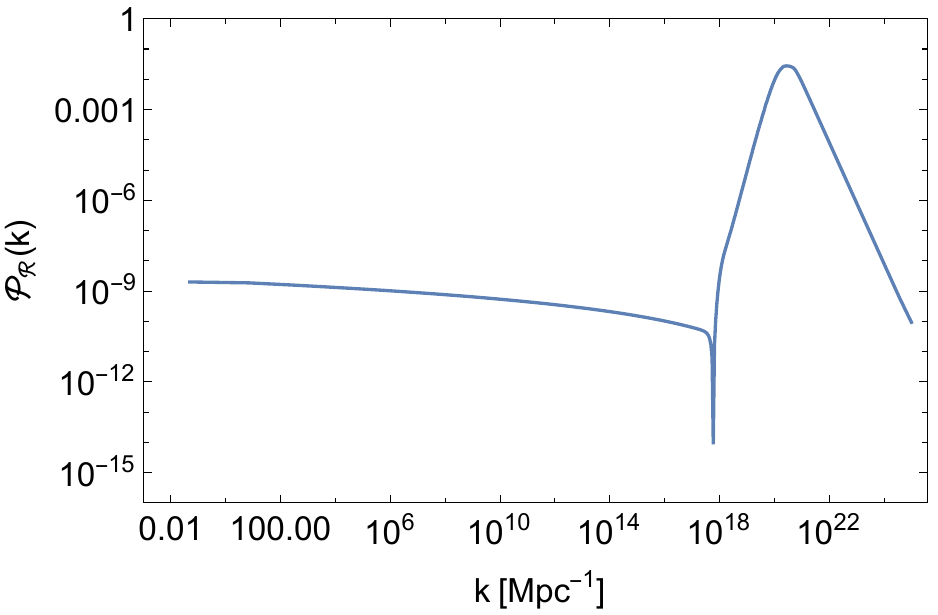}
\caption{The power spectrum of the comoving curvature perturbations for the 
model (\ref{fphi}) and potential depicted in Figure \ref{FigVSR}.
The power spectrum has a peak at large wavenumbers  triggering the production of mini PBHs during kination regime.
} \label{FigPSkin}
\end{figure}

After the numerical computation of the Mukhanov--Sasaki equation the  ${\cal P_ R}$  at all  scales is obtained, see Figure \ref{FigPSkin}. 

The position of the peak determines the PBH mass, but  the knowledge of the duration of kination and PBH domination phase is also required. According to the discussion in Section \ref{S5.2} the peak of the power spectrum is at the wavenumber,
\begin{align}
k_\text{peak}=k_\text{rh}\, e^{2\Delta N_\text{dom} +N_\text{eMD}/2}
\end{align}
where $k_\text{rh}\approx 2 \times 10^7 T_\text{rh}$ GeV$^{-1}$Mpc$^{-1}$.
Inflation ends at the wavenumber\linebreak $k_\text{end} = k_* e^{N_*} {H_\text{end}/H_*} $. 

A power spectrum with a peak at large wavenumbers, $k
\gg k_*$, is welcome because it is not spoiling the $n_s$ and $\alpha_s$ values measured at  $k_*$. 
Notwithstanding, shifting the peak at large wavenumbers does not render the ${\cal P_R}(k)$ free from constraints.
 Hawking radiation might affect BBN and CMB observables and upper bounds on the ${\cal P_R}(k)$ at large $k$-bands.
Therefore, one has to be careful with the width and the tail of the power spectrum peak since PBHs with a distribution of masses might be generated.
 The population of (non-monochromatic) PBHs with mass  $M_\text{PBH}\gtrsim 10^9$ g, if not zero, has to be small enough.
The stringent constraint is at the mass $M_\text{PBH} =2.5\times 10^{13}$~g, 
where the variance of the density perturbations has to satisfy  \cite{Dalianis:2018ymb},
\begin{align} \label{KB}
\sigma \, \lesssim \, 0.032 \, 
\left(\frac{\delta_c}{0.375}\right) \,.
\end{align}

This bound is easily satisfied in our scenario that we demand a curvature  power spectrum peak that generates ultra-light PBHs with mass $M_\text{PBH}\ll 10^{9}$~g, Figure \ref{FigPSkin}, but this is not always true for wide power spectra that generate a distribution of  PBH masses, in particular during an eMD era. For 
a wide curvature power spectrum with a peak close to the maximum $k$ values 
the growth of ${\cal P_R}(k)$ \cite{Byrnes:2018txb} together with the reheating temperature~\cite{Dalianis:2018ymb}  have to be checked.

From ${\cal P_ R}(k)$ we compute the $\beta$, $T_\text{rh}$ and $\Omega_\text{rem}$  as described in previous sections, neglecting possible impacts on the power spectrum from non-Gaussianities \cite{Franciolini:2018vbk, Passaglia:2018ixg, Atal:2018neu, DeLuca:2019qsy} and quantum diffusion effects \cite{Pattison:2017mbe, Biagetti:2018pjj, Ezquiaga:2018gbw, Cruces:2018cvq}.

\subsection{Inflection Point}
Inflaton potentials with an inflection point can produce a  power spectrum of curvature perturbations ${\cal P_R}(k)$ with large hierarchies. 
At the region of the inflection point the amplitude can be amplified by many orders of magnitude due to the acceleration and deceleration of the inflaton~\cite{Garcia-Bellido:2017mdw, Motohashi:2017kbs, Ballesteros:2017fsr}.
Such a model may arise from the $\alpha$-attractors models~\cite{Kallosh:2013yoa}
as described in Ref. \cite{Dalianis:2018frf, Dalianis:2019asr, Iacconi:2021ltm}. Successful CMB observables and a significant PBH population can be generated  
if the power spectrum peak is positioned at large  $k$.

Furthermore, we note that peaks in the power spectrum can also be related to the existence of one or more sharp drops (steps) in the inflaton potential \cite{Kefala:2020xsx, Dalianis:2021iig}. One could replace the inflection point with one or more sharp steps sufficiently close to each other and an analogous results can be attained.

The effective Lagrangian for the inflaton field $\varphi$ in $\alpha$-attractors reads
\begin{equation} \label{la11}
e^{-1}{\cal L}= \frac{1}{2}R-\frac{1}{2}
\Big(\partial_\mu \varphi\Big)^2-f^2\Big(\tanh\frac{\varphi}{\sqrt{6\alpha}}\Big)\,,
\end{equation}
where Re$\Phi=\phi=\sqrt{3}\tanh({\varphi}/{\sqrt{6\alpha}})$ is a chiral superfield and we took $M_\text{Pl}=1$. 
Polynomial, trigonometric and exponential forms for the function $f(\phi)$ can feature   an inflection point plateau sufficient to generate a significant dark matter abundance in accordance with the observational constraints  \cite{Dalianis:2018frf}. 
An explicit  working example is a combination of exponentials  of the form, 
\begin{align} \label{fphi}
f(\phi/ \sqrt{3}) =f_0\, \left( c_0 + c_1 e^{\lambda_1 \phi/\sqrt{3}} + c_2 e^{\lambda_2 (\phi-\phi_\text{P})^2/3}\right)
\end{align}
 that generates the potential $V(\varphi)= |f(\phi/\sqrt{3})|^2$.
In the above expressions we  have taken $\alpha=1$, but other choices also work. The parameter $f_0$ is  redundant in the sense that it can be  absorbed by  $c_0$, $c_1$ and $c_2$, but it is kept for numeric convenience.

For $\phi \rightarrow \sqrt{3}$ the potential drives the cosmic inflation and the CMB normalization gives the first  constraint for the parameters. Furthermore, 
the $n_s$ and $dn_s/d\ln k$ values have to be in accordance with the CMB data, given by Planck 2018 collaboration \cite{Akrami:2018odb} as well as the recent bounds by BICEP \cite{BICEP:2021xfz}.
Requiring PBH production with the right abundance  chooses specific values for the parameters that  shape the inflection point plateau.
In the following we also demand the value of the runaway potential at the field value today to be slightly above zero in order to identify it as the dark energy.

As commented before, the precise determination of the parameter values is a subtle numerical process.
The $M_\text{PBH}$ and the efold number values $N_*$, $\Delta N_\text{dom}$, $N_\text{KIN}$ select the $k$-position of the ${\cal P_R}(k)$ peak.
The required $\beta$ value determines the amplitude of the ${\cal P_R}(k)$ peak which is found after a careful and precise selection of the potential parameters due to the exponential sensitivity of $\beta$ to the  amplitude.
 In our scenario particular attention has to be paid at the number of efolds $N_*$ that has to be in agreement with the postinflationary cosmic evolution that involves a kination and a PBH domination phase.

\section{From Inflation to Dark Energy}\label{S7}

A runaway inflaton has interesting implications not only for the early but, also, for the late universe cosmology. 
Regarding the early universe, the inflaton does not decay oscillating about a vacuum and  a period of  kinetic energy domination of the scalar field can be realized. 
It has distinct  early universe phenomenology due to the stiff  equation of state $w_\text{}\sim 1$ and the expansion rate is reduced. The most notable effect is that the tensor perturbations attain a spectrum with more power at small scales. 
Furthermore, the fact that a tiny trace of the inflaton field potential energy density might remain non-zero until the present-day universe draws the attention to these inflationary models that 
can play the role of quintessence.
In the framework of $\alpha$-attractors kination models have been constructed in \cite{Dimopoulos:2017zvq, Dimopoulos:2017tud, Akrami:2017cir}.

 After inflation the scalar field  rolls down the runaway potential until it freezes due to Hubble friction  at some value $\varphi_\text{F}$. The residual potential energy at the frozen value acts temporarily like a cosmological constant and must be compatible with early and late time cosmological observations, $ V(\varphi_\text{F}) \lesssim 10^{-120} M^4_\text{Pl}$.
  When the upper bound is saturated, the runaway inflaton field is identified as the quintessence field that drives the accelerated expansion today. 
The  inflation runaway potential originating from the theory (\ref{fphi}) can drive late time inflation after a proper choice of parameters.
We ask for $V(\phi)\rightarrow 0$ as $\varphi \rightarrow-\infty$, or equivalently $\phi \rightarrow -\sqrt{3}$ in order that the potential is positive-definite.
At the value $\varphi_\text{F}$ the  potential is  flat enough and with small enough slow-roll parameters to implement a wCDM   quintessence model. 
The hierarchy of energy scales between  $\alpha$-attractors inflation and the present dark energy
implies a tuning of the potential energy value at $\varphi_\text{F}$,
\begin{align} \label{third}
\frac{\rho_\text{inf}}{\rho(t_0)} \simeq \frac{V(\varphi\gg 1)}{V(\varphi_\text{F})}\sim \frac{e^{2\lambda_1}}{e^{-2\lambda_1}} \sim 10^{108}\,,
\end{align}
that is $\lambda_1 \sim {108}\ln(10)/4 \sim 62$.  This condition is a third constraint to the parameters of the potential, together with  the CMB normalization  and the positivity of the potential energy density. A working example is given by the set of parameter values $c_0= -8.7 \times 10^{-27}$, $c_1=0.1045$, $c_2 = -4\times 10^{25}$, $\lambda_1= 62.2$ and $\lambda_2 = -4430.97$, $f_0^2=3.115\times 10^{-62} M^4_\text{Pl}$, $\phi_\text{P}=0.995 M_\text{Pl}$ and initial field value $\varphi_\text{CMB}=11.8 M_\text{Pl}$. Details can be found in Ref. \cite{Dalianis:2019asr}.

\begin{figure}[!htbp]
 \hspace{-10pt}   \includegraphics[width=0.513 \linewidth]{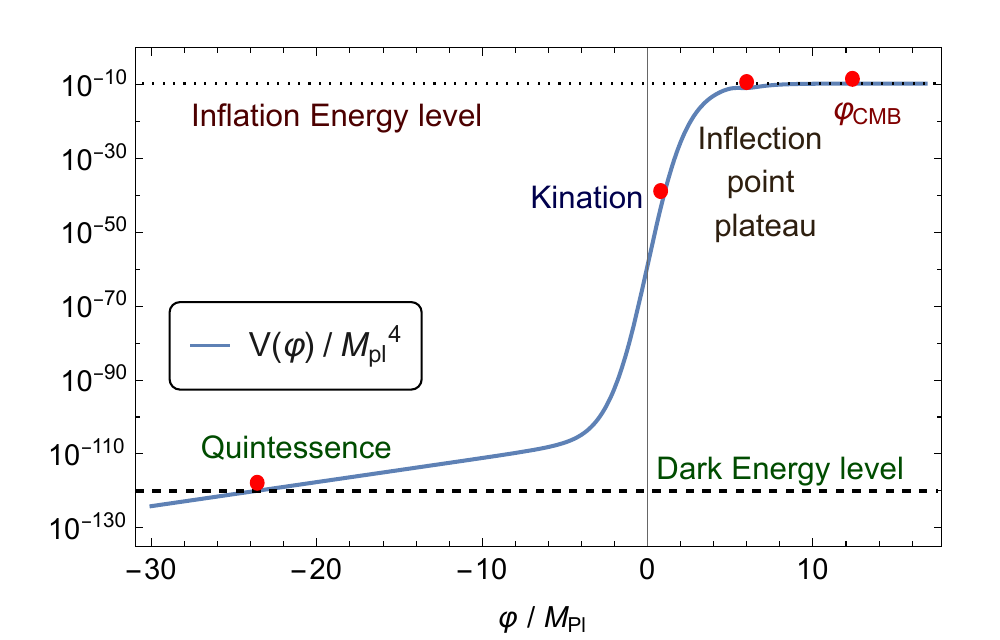}
  \includegraphics[width=0.49\linewidth]{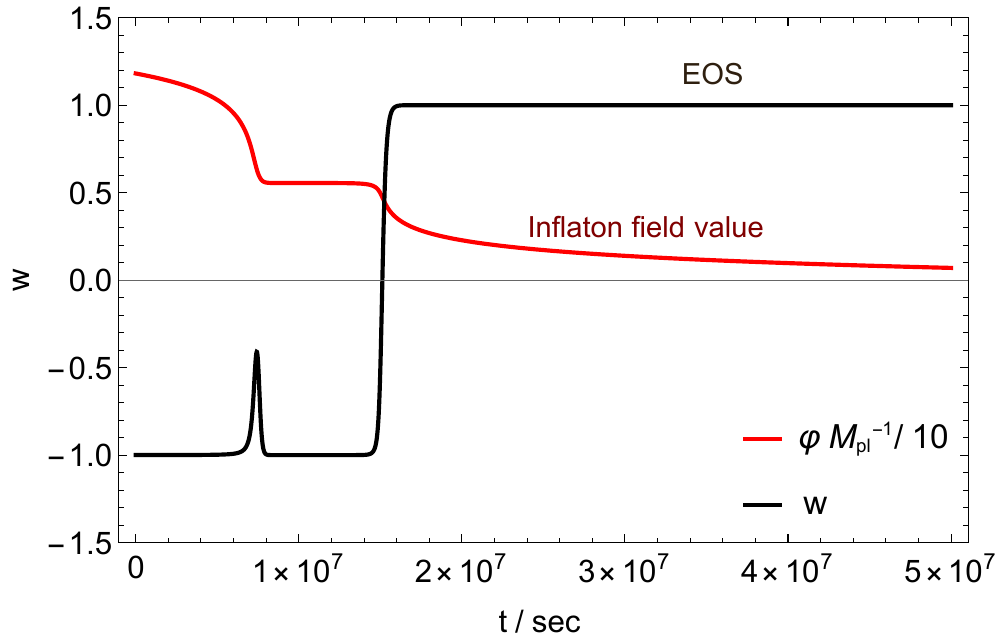}
\caption{In the \textbf{left panel} 
the unified inflaton-quintessence potential is  shown in logarithmic scale. The critical stages of reheating, the  PBH production and evaporation are highlighted.
In the \textbf{right panel} the equation of state (EOS) parameter and the inflaton field value are depicted making manifest the the presence of an infection point plateau and the postinflationary runaway phase.
} \label{FigSRw}
\end{figure}

\begin{figure}[!htbp] 
\includegraphics[width = 0.68\textwidth]{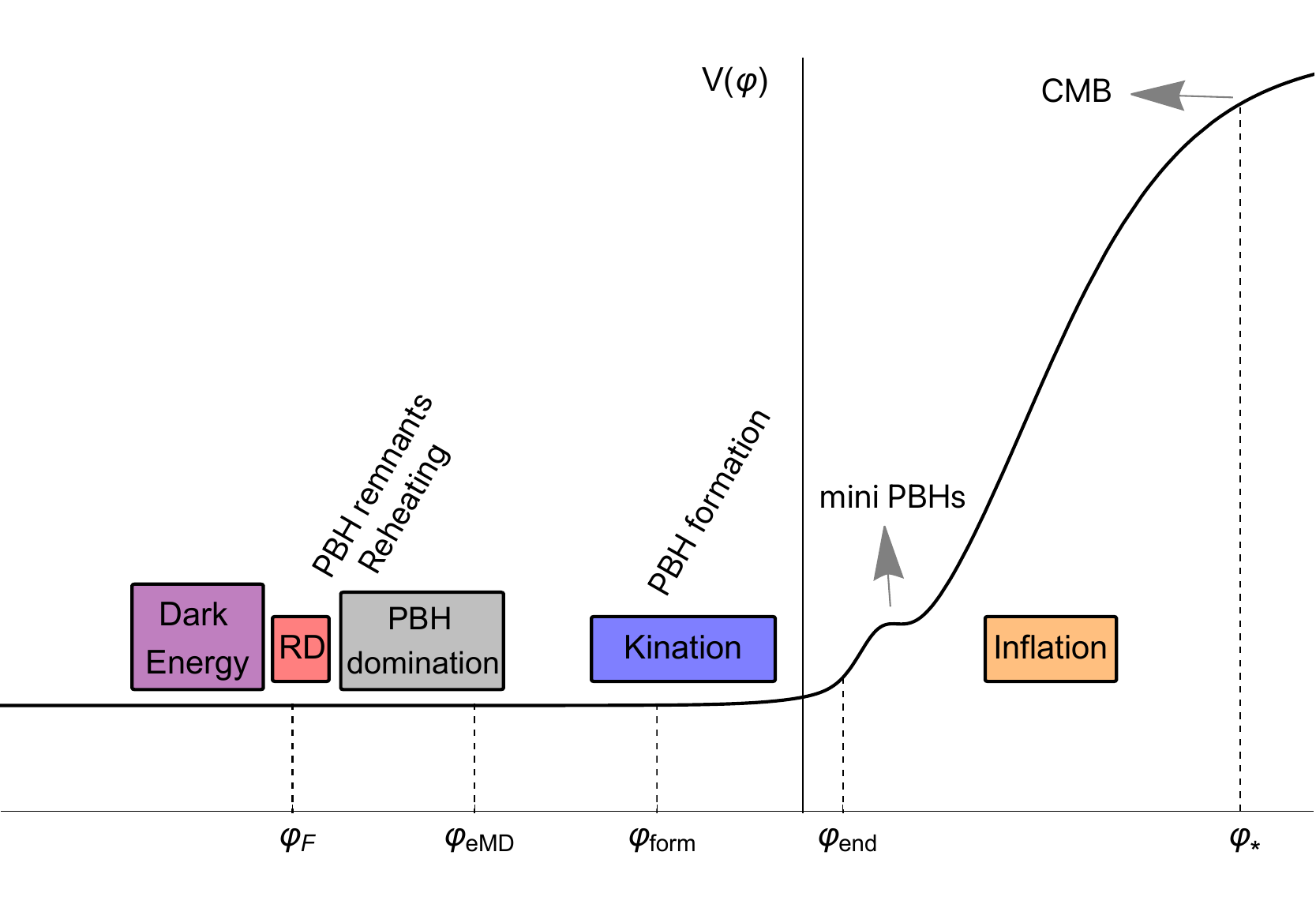}
\caption{A schematic illustration of the runaway inflationary model and the cosmic phases realized with respect to the inflaton field trajectory.} \label{FigCart}
\end{figure}

In summary, at the field values $\varphi_\text{CMB}$ and $\varphi_\text{P}$  inflation produces the seeds of CMB  anisotropies and PBHs, respectively, and at $\phi_\text{end}$ inflation ends and a kination stage commends. Later PBH form, dominate the energy density of the universe and evaporate reheating the universe.
At the field  value $\varphi_\text{F}$ the field freezes and its potential energy plays the role of the  dark energy in the universe. 
It is remarkable that the  simple theory (\ref{fphi}) 
provides a complete example that can explain the main cosmological observations, for appropriate though highly tuned values of the parameters.
Next, we will give an approximate  quantitative description of  the aforementioned  stages of this cosmological model, that we also illustrate in Figures \ref{FigSRw} and \ref{FigCart}.

\subsection*{The Evolution of the Runaway Inflaton}

Let us work out some  simple analytic expressions and describe  approximately but quantitatively the post-inflationary evolution of the field $\varphi$. 
After inflation $\varphi$ fast-rolls the potential and a stage of kination domination takes over  where $\dot{\varphi}^2/2 \gg V(\varphi)$.
For negligible potential energy the evolution of the scalar field is given by the system of equations
\begin{align} \label{EoM1}
& 6H^2 M^2_\text{Pl} \approx \dot{\varphi}^2 \nonumber \\ 
&\ddot{\varphi} +3H \dot{\varphi} \approx 0
\end{align}

From the Friedmann equation we obtain $H M_\text{Pl}\approx \pm \dot{\varphi}/\sqrt{6}$,
 where the minus is for negative field velocity.
After an integration  the evolution of the field value is obtained,
\begin{align}\label{velKin}
\varphi-\varphi_\text{init} \approx  \pm \sqrt{\frac{2}{3}} M_\text{Pl} \ln \left[1 \pm \frac{\dot{\varphi}_\text{init}}{M_\text{Pl}}\sqrt{\frac{3}{2}} (t-t_\text{init}) \right]\,.
\end{align}
where $t_\text{init}$ can be defined as the moment that $\dot{\varphi}^2$ dominates the energy density. 
The\linebreak initial velocity is found from the Friedmann equation and the Hubble parameter,\linebreak 
 \mbox{$\dot{\varphi}_\text{init}\approx \pm M_\text{Pl} H_\text{init} \sqrt{6} \approx \pm M_\text{Pl} t_\text{init}^{-1} \sqrt{2/3}$}. 
At the end of inflation $t_\text{end}$ it is $V=\dot{\varphi}^2$, 
but soon afterwards kination regime takes over, see Figure \ref{FigSRw}. We
can approximate $t_\text{init}$ with the moment inflation ends, 

\begin{align} \label{phiend1}
\varphi(t)-\varphi_\text{end} \approx -\sqrt{\frac{2}{3}} M_\text{Pl} \ln\left( \frac{t}{t_\text{end}}\right)\, \quad \text{for} \quad t_\text{end}< t< t_\text{rh}\,,
\end{align} 
where we also considered negative initial field  velocity. 
At the moment $t_\text{form}$ PBHs form and later at $t_\text{evap}$ they evaporate.

If there is no PBH domination phase,
the universe becomes radiation dominated at  the moment $t_\text{rh}$ and kination regime ends,
where $t_\text{end}$ and $t_\text{rh}$ define the duration of the kination phase. 
Until reheating it is  approximately  $a \propto t^{1/3}$ 
the reheating moment is found to be $t_\text{rh}=\left(\Omega_\text{rad}(t_\text{evap})\right)^{-3/2} t_\text{evap}$,
where  $\Omega_\text{rad}(t_\text{evap})=(3/2)\gamma^2 \beta M_\text{PBH}^2/m^2_\text{Pl}$, given by \mbox{Equation (\ref{KinEvap})}.
Utilizing the above  results,  the 
field value at the reheating moment, \mbox{$\varphi(t_\text{rh})\equiv \varphi_\text{rh}$}, is found.
 After reheating, we can follow the same steps and assume that the scalar field keeps running away its potential  
 with the Hubble parameter given by  \mbox{$H\approx 1/(2t)$} (radiation domination). Thus, in the reheated universe the scalar field evolves~as 
\begin{align} \label{phireh2}
\varphi(t)\approx  \varphi_\text{rh}  -\frac{2 M_\text{Pl}}{\sqrt{3}} \left( 1-\sqrt{\frac{t_\text{rh}}{t}}\right)\,.
\end{align}
where we approximated   $H^2(t_\text{rh})\approx \dot{\varphi}^2_\text{rh}/{3 M^2_\text{Pl}}$ at the reheating time.
Notice that the field is not rolling fast the runaway potential but slows down  and at late times $t  \gg t_\text{rh}$ the field freezes having covered a $2 M_\text{Pl}/\sqrt{3}$ distance in field space  from the value $\varphi_\text{rh}$.

Next we add a PBH domination phase, that is necessary for viable cosmology,  according to the discussion in Section \ref{S5}.

\subsubsection*{PBH Domination Phase}
In the viable case that a PBH domination phase takes place, the kination phase ceases earlier, at the time $t_\text{eMD}$, where eMD stands for early matter domination. There is  the following  hierarchy of  moments,
\begin{align}
t_\text{end} <t_\text{form}<t_\text{eMD} <t_\text{evap}\equiv t_\text{rh} <t_\text{F}\,.
\end{align}

PBHs form with a nearly instantaneous collapse 
at the time 
\begin{align}
t_\text{form}=\frac{2}{3}\frac{M_\text{hor}}{m^2_\text{Pl}} = \frac{2}{3}\frac{M_\text{PBH}/\gamma}{m^2_\text{Pl}}.
\end{align}

Since $\rho_\text{PBH}/\rho_\varphi \propto a^{3}$ PBHs dominate the energy density of the universe if their evaporation time is larger than $t_\text{eMD}$,
\begin{align}
t_\text{eMD}= \frac{t_\text{form}}{\gamma \beta},
\end{align}
 which is found after equating PBH energy density with that of the runaway inflaton. 
For $t>t_\text{eMD}$ we can approximately find the scalar field value to be, 
\begin{align}
\varphi(t) &  \approx \varphi_\text{eMD} -\frac{M_\text{Pl}}{\sqrt{3}} \left( 1-{\frac{t_\text{eMD}}{t}}\right)  
\end{align}

If $t_\text{eMD}\ll t_\text{evap}$ the value that the field freezes is given roughly by
\begin{align}
\varphi_\text{F} \approx \varphi_\text{end}-\sqrt{\frac{2}{3}}
\left(\frac{\sqrt{2}}{2}-\ln (\gamma\beta) + \ln\left(\frac{t_\text{form}}{t_\text{end}}\right)\right)M_\text{Pl}\,.
\end{align}
where $\gamma \beta =\Omega_\text{PBH}(t_\text{form})$. 
  If $t_\text{eMD}\sim t_\text{evap}$ the evolution during the radiation era will add a displacement of size $2 M_\text{Pl}/\sqrt{3}$.

Interestingly enough 
we can write, for a PBH domination phase, 
 the value that inflaton freezes as a function of  $r_*$, $\beta$ and PBH mass as follows,
\begin{align} \label{fFeMDnew}
\varphi_{F}\sim \varphi_\text{end} - \sqrt{\frac{2}{3}} 
\left[27.7 + \ln\left(\frac{M_\text{PBH}/\gamma}{10^5 \text{g}}\right)-\ln\left(\frac{\gamma \beta}{10^{-8}}\right) + \ln \sqrt{r_*}  \right]M_\text{Pl}\,.
\end{align} 

We mention that the above expressions are rough approximations and should be seen only as indicative since the potential has been neglected and instant transitions have been assumed. 
Evidently,  the exact value of  $\varphi_\text{F}$ is found only after the Klein--Gordon and Friedman equations are solved  numerically.

\section{Conclusions and Discussion}\label{S8}

In this work, we investigated the reheating of runaway 
inflation models via the evaporation of primordial black holes.
A runaway inflaton potential has no minimum and the inflaton field does not decay but gradually loses potential energy. 
Since the standard mechanism of reheating cannot be realized in this type of models, another 
source of entropy production has to be introduced.
Black holes are objects that exist in the present-day universe having a quite significant population and it is plausible to presume that they might have existed in the early universe as well. Mini black holes are motivated candidates because the early universe, although homogeneous at large scales, might be characterized by strong  small scale fluctuations. 
If these fluctuations have size less than about $k^{-1}\sim 10^{-19}$~Mpc and density contrast that exceeds a threshold value, gravitational collapse takes place and primordial black holes with an ultra light mass are formed.
 These mini black holes are ephemeral and promptly transform their  entire mass into thermal degrees of freedom. Black hole evaporation is therefore an alternative reheating mechanism that can render, the otherwise problematic, runaway inflation models cosmologically viable.
 
 Primordial black holes with mass $M_\text{PBH}<10^9$ g have to be generated with a significant population in the early universe in order to implement a successful reheating. The abundance of the PBHs is determined by the parameter $\beta$ that measures the fraction of the universe energy density that collapsed.
 At the same time with the PBH generation a substantial amount of gravitational radiation is produced, called induced or secondary GWs.  Contrary to the ephemeral mini PBHs, the associated induced GWs propagate in the early and late universe and have significant observational implications. 
First, their energy density contributes to the expansion rate and impacts BBN and CMB. Second, the secondary GWs are in principle detectable by current and near future gravitational wave~detectors. 

The size of the $\beta$ value plays a critical role and determines the details of the cosmological evolution. Small $\beta$ values imply a small population of PBHs and an extended kination phase. Kination domination ends due to the radiation produced by PBH evaporation at a temperature $T_\text{rh}\propto \beta^{3/4}$. Although reheating temperatures safely larger than ${\cal O}(10)$ MeV can be realized, the extended kination scenario is ruled out. The energy density of GWs, of both primary and secondary, becomes enhanced and alters  BBN and CMB. Noteworthy, the stringent constraint on $\beta$ comes from the secondary GWs. The $\beta$ value has to be large enough, $\beta\gtrsim 10^{-18}-10^{-10}$, so that PBHs dominate the energy density of the early universe, Equation (\ref{masterbound}).  A PBH domination compensates the relative increase of the GW energy density that occurs during kination domination. 
 In the present-day universe the induced GWs have a non-negligible amplitude but too large frequencies, which generally lay outside the sensitivity band of the current or near future detectors.  A PBH domination phase comes with an isocurvature perturbation which, in turn, induces an extra component of secondary GWs   \cite{Papanikolaou:2020qtd} that constrains the maximum $\beta$ value  \cite{Domenech:2020ssp}.  
 The final allowed parameter space ($M_\text{PBH}, \beta$) is summarized in Figure \ref{FigAll}. It is exciting that the allowed parameter space of our scenario can shrink further from data coming from GW experiments as well as from cosmological precision measurements of the effective relativistic  degrees of freedom $\Delta N_{\nu,\text{eff}}$ during BBN or photon decoupling.

A very interesting implication of reheating runaway inflation models via PBH evaporation is that, on the one hand,  the inflaton field has a non-zero 
potential energy in the present-day universe and, on the other, PBH evaporation might leave a remnant mass behind. The former form of energy can play the role of  dark energy. The latter can constitute the dark matter, or part of it, in the galaxies. The mass of the remnant is expected to depend on the unknown physics that operates at the Planck energy scale or somewhere close to that scale.
PBH remnants have a significant cosmological abundance for a particular remnant mass  $M_\text{rem}$ which is proportional to $M_\text{PBH}^{5/2}$, see Equation  (\ref{fremMD}), and lies in the range between $10^{-15}m_\text{Pl}$ and $10^7m_\text{Pl}$. 
Due to the near one-to-one cosmological correspondence between $M_\text{rem}$ and $M_\text{PBH}$ a potential discovery of those exotic remnants will connect us, among others, with the early universe and PBHs.  
The reheating temperature of the universe, which also depends on the  PBH mass as $M_\text{PBH}^{-3/2}$, can range from few MeV up to $10^{10}$ GeV. Let us note that the reheating temperature has important implications for other essential  processes such as baryogenesis and dark matter production, topics which are outside the scope of this work. 

Last, we engaged in an explicit inflationary model built in the framework of $\alpha$-attractors \cite{Dalianis:2019asr} to implement this cosmological scenario. Mini PBHs are totally compatible with the inflationary observables, in particular the  spectral index value,  since the power spectrum has to be enhanced at very large wavenumbers far away from the CMB scale.
 It is remarkable that the minimal theory (\ref{fphi})  of a single field with a non-canonical kinetic term and without the need of extra interactions,
 can explain
the basic cosmological observations.  It gives cosmic inflation, reheats the universe via the production and evaporation of mini black holes, 
predicts the presence of dark matter if the black holes leave a remnant behind, and drives the late acceleration of the universe via the residual vacuum energy of the very same scalar field. 
Let us comment that, although extremely economic, 
  we advocate the presence of new physics that could give rise to such a type of  quintessential inflationary field with inflection point or strong features in its inflationary  trajectory.

\appendix

\section{Gravitational and Instant (P)reheating}\label{A1}

Let us  briefly describe two other non-conventional reheating mechanisms:  gravitational reheating and instant (p)reheating. The first mechanism is inefficient regarding the current observational constraints, while the second one depends on the coupling of the inflaton to another scalar field. This is why we regard them beyond the main line of the~paper.

\subsection{Gravitational Reheating}

Gravitational particle production \cite{Ford:1986sy, Chun:2009yu, Dimopoulos:2017zvq} is an inevitable but feeble reheating mechanism. Particle production of all light fields (with masses less than the Hubble scale) takes place due to the change of the spacetime metric at the end of inflation.
This is essentially Hawking radiation in de Sitter space, which generates a radiation bath of temperature $T_H = H/2\pi$. We assume that the energy density at the end of inflation is dominated by the kinetic energy of the scalar field (kination domination).
The radiation density produced by gravitational reheating at the end of inflation is

\begin{equation}
\rho_{\textrm{rad}}^{\textrm{end}} = q \frac{\pi^2}{30} g_{*}^{\text{end}} \left( \frac{H_{\textrm{end}}}{2\pi} \right)^4
\label{rho_r_end}
\end{equation}
where $g_{*}^{\textrm{end}}$ is the effective number of relativistic degrees of freedom at the energy scale of inflation and $q \sim 1$ is an efficiency factor.

Using Equation (\ref{rho_r_end}), we have for the radiation density parameter at the end of inflation:
\begin{equation}
\Omega_{\textrm{rad}}^{\textrm{end}} \equiv \frac{\rho_{\textrm{rad}}^{\textrm{end}}}{\rho_{\textrm{tot}}^{\textrm{end}}} = \frac{q g_{*}^{\textrm{end}}}{1440\pi^2} \left( \frac{H_{\textrm{end}}}{M_\text{Pl}} \right)^2
\label{Omega_r_end}
\end{equation}
where we have used that $\rho_{\textrm{tot}}= 3H^2 M_\text{Pl}^2$.
The universe is reheated when radiation takes over and dominates the kinetic density of
the scalar field. This definitely happens, because the energy density of kination is red-shifted away much faster than that of radiation as $\rho_{\textrm{KIN}} \propto  a^{-6}$ and  $\rho_{\textrm{rad}} \propto a^{-4}$.
Using that $a \propto t^{1/3}$ during kination, it is easy to show that the time when reheating occurs is
\begin{equation}
t_{\textrm{rh}} = (\Omega_{\textrm{rad}}^{\textrm{end}})^{-3/2} t_{\textrm{end}} .
\label{trh}
\end{equation}

As $\Omega_{\textrm{rad}} =\rho_{\textrm{rad}}/\rho_{\textrm{KIN}}$ during kination, we can easily find that $\rho_{\textrm{KIN}}^{\textrm{rh}} =(\Omega_{\textrm{rad}}^{\textrm{end}})^3 \rho_{\phi}^{\textrm{end}}$. Thus, considering that radiation is thermalized by the time it comes to dominate the Universe, we find that the reheating temperature is
\begin{equation}
T_{\textrm{rh}} = \left[ \frac{30}{\pi^2 g_{*}^{\textrm{rh}}} (\Omega_{\textrm{rad}}^{\textrm{end}})^3 \rho_{\varphi}^{\textrm{end}} \right]^{1/4},
\label{Trh1}
\end{equation}
which, using Equation (\ref{Omega_r_end}), can be written as
\begin{equation}
T_{\textrm{rh}} = \frac{q^{3/4}}{24\pi^2} \left( \frac{g_{*}^{\textrm{end}}}{g_{*}^{\textrm{rh}}} \right)^{1/4}   \left(\frac{g_{*}^{\textrm{end}}}{10}  \right)^{1/2}  \frac{H_{\textrm{end}}^2}{M_{\text{Pl}}}.
\label{Trh2}
\end{equation}

However, we can easily see that this reheating temperature does not obey the BBN constraint (\ref{TGW}).
For example, if we set $H_{\textrm{end}} \sim 10^{12}$ GeV, $g_{*}^{\textrm{end}} = {\cal O} (100)$ and \mbox{$g_{*}^{\textrm{rh}} = 10.75$}, Equation (\ref{Trh2}) gives $T_\text{rh}\approx 10^4$ GeV, while Equation (\ref{TGW}) demands that $T_\text{rh} \gtrsim 10^{7}$ GeV.
Consequently, gravitational reheating is not efficient and we should rely on other mechanisms.

\subsection{Instant (P)reheating}

Instant preheating can work both in usual inflationary models where $V(\phi)$ has a minimum and in quintessential models such as those that we examine in this paper \cite{Felder:1998vq, Dimopoulos:2017tud}.
The basic assumption is that the inflaton field $\phi$ is coupled to another scalar field $\chi$ which is also coupled to a fermion field $\psi$. The interaction Lagrangian density is
\begin{equation}
{\cal L}_{\textrm{inter}} = -\frac{1}{2} g^2 (\phi-\phi_0)^2 \chi^2 - h \chi \psi \bar{\psi},
\label{L_inter}
\end{equation} 
where $g$ and $h$ are perturbative coupling constants.
In order for particle production to occur, we must have \cite{Dimopoulos:2017tud}
\begin{equation}
|\dot{\phi}| > g (\phi-\phi_0)^2,
\end{equation}
which gives the range for $\phi$:
\begin{equation}
\phi_0 - \sqrt{\frac{|\dot{\phi}|}{g}} \leq \phi \leq \phi_0 + \sqrt{\frac{|\dot{\phi}|}{g}}.
\label{phi_range}
\end{equation}

The density of the produced $\chi$ particles can be shown to be \cite{Felder:1998vq}

\begin{equation}
\rho_{\chi}^\text{InP}= \frac{g^{5/2}|\dot{\phi}^\text{InP}|^{3/2}\phi^\text{InP}}{8\pi^3}.
\label{rho_chi}
\end{equation}
where the superscript ``InP” denotes instant preheating values. In fact, we expect only the particles produced near the end of the particle production window in (\ref{phi_range}) to contribute significantly to $\rho_{\chi}^\text{InP}$ \cite{Dimopoulos:2017tud}. Thus, taking $\phi_0 = 0$, we can set $\phi^\text{InP} = \sqrt{\frac{|\dot{\phi}^\text{InP}|}{g}}$. Consequently Equation (\ref{rho_chi}) can be written in the simpler form
\begin{equation}
\rho_{\text{rad}}^\text{InP}= \rho_{\chi}^\text{InP}= \frac{g^2 (\dot{\phi}^\text{InP})^2}{8\pi^3},
\label{rho_x}
\end{equation}
where we have considered that $\chi$-particles decay to radiation instantaneously.
For the reheating temperature in this model we have
$\rho_{\text{rad}}^{\text{rh}} = \rho_{\phi}^{\text{rh}} = \rho_{\phi}^{\text{InP}} (\Omega_\text{rad}^\text{InP})^3 $.
Thus,
\begin{equation}
T_{\text{rh}} = \left[ \frac{30}{\pi^2 g_{*}^{\text{rh}}} \rho_{\phi}^{\text{InP}} (\Omega_\text{rad}^\text{InP})^3   \right]^{1/4},
\end{equation}
where $\Omega_\text{rad}^\text{InP}$ can is found to be $\Omega_\text{rad}^\text{InP} = {g^2}/(4\pi^3)$.
Consequently  in instant preheating $T_{\text{rh}}$ depends strongly on the value of the coupling constant $g$ and takes larger values as we increase $g$. However, the region of $g$ so that all cosmological constraints are satisfied is given approximately by \cite{Dimopoulos:2017tud}
\begin{equation}
10^{-4} \lesssim g \lesssim 10^{-2}.
\end{equation}

\section*{Acknowledgments}

\noindent 
We would like thank N. Tetradis for discussions. This work is supported by the Hellenic Foundation for Research and Innovation
(H.F.R.I.) under the “First Call for H.F.R.I. Research Projects to support Faculty members and
Researchers and the procurement of high-cost research equipment grant” (Project Number: 824).

\noindent


\end{document}